# Clustering High Dimensional Dynamic Data Streams

Vladimir Braverman[*]  
Johns Hopkins University

Gereon Frahling[†]  
Linguee GmbH

Harry Lang[‡]  
Johns Hopkins University

Christian Sohler[§]  
TU Dortmund

Lin F. Yang[¶]  
Johns Hopkins University

June 12, 2017

## Abstract

We present data streaming algorithms for the $k$-median problem in high-dimensional dynamic geometric data streams, i.e. streams allowing both insertions and deletions of points from a discrete Euclidean space $\{1, 2, \ldots \Delta\}^d$. Our algorithms use $k\epsilon^{-2}\mathrm{poly}(d \log \Delta)$ space/time and maintain with high probability a small weighted set of points (a coreset) such that for every set of $k$ centers the cost of the coreset $(1 + \epsilon)$-approximates the cost of the streamed point set. We also provide algorithms that guarantee only positive weights in the coreset with additional logarithmic factors in the space and time complexities. We can use this positively-weighted coreset to compute a $(1 + \epsilon)$-approximation for the $k$-median problem by any efficient offline $k$-median algorithm. All previous algorithms for computing a $(1 + \epsilon)$-approximation for the $k$-median problem over dynamic data streams required space and time exponential in $d$. Our algorithms can be generalized to metric spaces of bounded doubling dimension.

---

[*]Email: `vova@cs.jhu.edu`. This material is based upon work supported is supported by the NSF Grants IIS-1447639, EAGER CCF- 1650041, and CAREER CCF-1652257.

[†]Email: `gereon@frahling.de`

[‡]Email: `hlang8@jhu.edu`. This material is based upon work supported by the Franco-American Fulbright Commission. H. Lang thanks INRIA (l'Institut national de recherche en informatique et en automatique) for hosting him during the writing of this paper.

[§]Email: `christian.sohler@tu-dortmund.de`. The author acknowledges the support of DFG within the Collaborative Research Center SFB 876, project A2.

[¶]Email: `lyang@pha.jhu.edu`. This material is based upon work supported by the NSF Grant IIS-1447639.

# 1 Introduction

The analysis of very large data sets is still a big challenge. Particularly, when we would like to obtain information from data sets that occur in the form of a data stream like, for example, streams of updates to a data base system, internet traffic and measurements of scientific experiments in astro- or particle physics (e.g. [LIY+15]). In such scenarios it is difficult and sometimes even impossible to store the data. Therefore, we need algorithms that process the data sequentially and maintain a summary of the data using space much smaller than the size of the stream. Such algorithms are often called streaming algorithms (for more introduction on streaming algorithms, please refer to [Mut05]).

One fundamental technique in data analysis is clustering. The idea is to group data into clusters such that data inside the same cluster is similar and data in different clusters is different. Center based clustering algorithms also provide for each cluster a cluster center, which may act as a representative of the cluster. Often data is represented as vectors in $\mathbb{R}^d$ and similarity between data points is often measured by the Euclidean distance. Clustering has many applications ranging from data compression to unsupervised learning.

In this paper we are interested in clustering problems over dynamic data streams, i.e. data streams that consist of updates, for example, to a database. Our stream consists of insert and delete operations of points from $\{1,\ldots,\Delta\}^d$. We assume that the stream is consistent, i.e. there are no deletions of points that are not in the point set and no insertions of points that are already in the point set. We consider the $k$-median clustering problem, which for a given a set of points $P \subseteq \mathbb{R}^d$ asks to compute a set $C$ of $k$ points that minimizes the sum of distances of the input points to their nearest points in $C$.

## 1.1 Our Results

We develop the first $(1+\epsilon)$-approximation algorithm for the $k$-median clustering problem in dynamic data streams that uses space polynomial in the dimension of the data. To our best knowledge, all previous algorithms required space exponentially in the dimension. Formally, our main theorem states,

**Theorem 1.1** (Main Theorem). *Fix $\epsilon \in (0, 1/2)$, positive integers $k$ and $\Delta$, Algorithm 1 makes a single pass over the dynamic streaming point set $P \subset [\Delta]^d$, outputs a weighted set $S$, such that with probability at least $0.99$, $S$ is an $\epsilon$-coreset for $k$-median of size $O\left(kd^4L^4/\epsilon^2\right)$, where $L = \log \Delta$. The algorithm uses $\tilde{O}\left(kd^7L^7/\epsilon^2\right)$ bits in the worst case, processes each update in time $\tilde{O}(dL^2)$ and outputs the coreset in time $\text{poly}(d, k, L, 1/\epsilon)$ after one pass of the stream.*

The theorem is restated in Theorem 3.6 and the proof is presented in Section 3.3. The coreset we constructed may contain negatively weighted points. Thus naïve offline algorithms do not apply directly to finding $k$-clustering solutions on the coreset. We also provide an alternative approach that output only non-negatively weighted coreset. The new algorithm is slightly more complicated. The space complexity and coreset size is slightly worse than the one with negative weights but still polynomial in $d$, $1/\epsilon$ and $\log \Delta$ and optimal in $k$ up to polylog$k$ factor.

**Theorem 1.2** (Alternative Results). *Fix $\epsilon \in (0, 1/2)$, positive integers $k$ and $\Delta$, Algorithm 6 makes a single pass over the streaming point set $P \subset [\Delta]^d$, outputs a weighted set $S$ with* non-negative weights *for each point, such that with probability at least $0.99$, $S$ is an $\epsilon$-coreset for $k$-median of size $\tilde{O}\left(kd^4L^4/\epsilon^2\right)$. The algorithm uses $\tilde{O}\left(kd^8L^8/\epsilon^2\right)$ bits in the worst case. For each update of* the input, the algorithm needs $\text{poly}(d, 1/\epsilon, L, \log k)$ time to process and outputs the coreset in time $\text{poly}(d, k, L, 1/\epsilon)$ after one pass of the stream.

The theorem is restated in Theorem 4.3 in Section 4 and the proof is presented therein. Both approaches can be easily extended to maintain a coreset for a general metric space.

## 1.2 Our Techniques

From a high level, both algorithms can be viewed as a combination of the ideas introduced by Frahling and Sohler [FS05] with the coreset construction by Chen [Che09].

To explain our high-level idea, we first summarize the idea of Chen [Che09]. In their construction, they first obtain a $(\alpha, \beta)$-bi-criterion solution. Namely find a set of at most $\alpha k$ centers such that the $k$-median cost to these $\alpha k$ centers is at most $\beta\text{OPT}$, where OPT is the optimal cost for a $k$-median solution. Around each of the $\alpha k$ points, they build logarithmically many concentric ring regions and sample points from these rings. Inside each ring, the distance from a point to its center is upper and lower bounded. Thus the contribution to the optimal cost from the points of this ring is lower bounded by the number of points times the inner diameter of the ring. To be more precise, their construction requires a partition of $\tilde{O}(\alpha k)$ sets of the original data points satisfying the following property: for the partition $P_1, P_2, \ldots, P_{k'}$, $\sum_i |P_i|\text{diam}(P_i) \lesssim \beta\text{OPT}$. They then sample a set of points from each part to estimate the cost of an arbitrary $k$-set from $[\Delta]^d$ up to $O(\epsilon|P_i|\text{diam}(P_i)/\beta)$ additive error. Combining the samples of the $k'$ parts, this gives an additive error of at most $\epsilon\text{OPT}$ and therefore an $\epsilon$-coreset.

The first difficulty in generalizing the construction of Chen to dynamic streams is that it depends on first computing approximate centers, which seems at first glance to require two passes. Surprisingly (since we would like to be polynomial in $d$), we can resolve this difficulty using a grid-based construction. The grid structure can be viewed as a $(2^d)$-ary tree of cells. The root level of the tree is a single cell containing the entire set of points. Going down a level through the tree, each parent cell is split evenly into $2^d$ subcells. Thus in total there are $\log_2 \Delta$ grid levels. Each cell of the finest level contains at most a single point.

Without using any information of a pre-computed $(\alpha, \beta)$-bi-criterion solution to the $k$-median problem, as it does in [Che09], our first idea (similar to the idea used in [Ind04]) is to use a randomly shifted grid (i.e. shift each coordinate of the cell of the root level by a random value $r$, where $r$ is uniformly chosen from $\{1, 2, \ldots \Delta\}$, and redefine the tree by splitting cells into $2^d$ subcells recursively). We show that with high probability, in each level, at most $\tilde{O}(k)$ cells are close to (or containing) a center of an optimal solution to the $k$-median. For the remaining cells, we show that each of them cannot contain too many points, since otherwise they would contribute too much to the cost of the optimal solution (since each point in these cells is far away from each of the optimal centers). We call the cells containing too many points in a level *heavy* cells. The immediate non-heavy children of the heavy cells form a partition of the entire point sets (i.e. the cells that are not heavy, but have heavy parents). Let $\mathcal{C}_1, \mathcal{C}_2, \ldots, \mathcal{C}_{k'}$ be these cells, and we can immediately show that $\sum_i |\mathcal{C}_i|\text{diam}(\mathcal{C}_i) \leq \beta\text{OPT}$ for some $\beta = O(d^{3/2})$. If we can identify the heavy cells (e.g. use heavy hitter algorithms), and sample points from their immediate non-heavy children in a dynamic stream, we will obtain a construction similar to Chen [Che09].

Our second idea allows us to significantly reduce space requirements and also allows us to do the sampling more easily. For each point $p$, the cells containing it form a path on the grid tree. We write each point as a telescope sum as the cell centers on the path of the point ( recall that the grids of each level are nested and the $c^0$ is the root of the tree). For example, let $c^0, c^1, \ldots, c^L$ be the cell centers of the path, where $c_L = p$, and define $\mathbf{0}$ to be the zero vector. Then $p = c^L - c^{L-1} + c^{L-1} - c^{L-2} \ldots + c^0 - \mathbf{0} + \mathbf{0}$. In this way, we can represent the distance from a point to a set of points as the distance of cell centers to that set of points. For example, let $Z \subset [\Delta]^d$ be a set of points, and $d(p, Z)$ be the distance from $p$ to the closest point in $Z$. Then $d(p, Z) = d(c^L, Z) - d(c^{L-1}, Z) + d(c^{L-1}, Z) - d(c^{L-2}, Z) + \ldots + d(\mathbf{0}, Z)$. Thus we can decompose the cost a set $Z$ into $L + 2$ levels: the cost in level $l \in [0, L]$ is $\sum_p d(c_p^l, Z) - d(c_p^{l-1}, Z)$, where $c_l^p$ is the center of the cell containing $p$ in level $l$ and the cost in the $(-1)$-st level is $|P|d(\mathbf{0}, Z)$, where $P$ is the entire points set. Since $|d(c_p^l, Z) - d(c_p^{l-1}, Z)|$ is bounded by the cell diameter of the level, we can sample points from the non-heavy cells of the entire level, and guarantee that the cost of that level is well-approximated. Notice that (a) we do not need to sample $\tilde{O}(k)$ points from every part of the partition, thus we save a $k$ factor on the space and (b) we do not need to sample the actual points, but only an estimation of the number of points in each cell, thus the sampling becomes much easier (there is no need to store the sampled points).

In the above construction, we are able to obtain a coreset, but the weights can be negative due the the telescope sum. It is not easy find an offline $k$-median algorithm to output the solutions from a negatively-weighted coreset. To remove the negative weights, we need to adjust the weights of cells. But the cells with a small number of points (compared to the heavy cells) are problematic – the sampling-based empirical estimations of the number of points in them has too much error to be adjusted.

In our second construction, we are able to remove all the negative weights. The major difference is that we introduce a cut-off on the telescope sum. For example, $d(p, Z) = d(c_p^L, Z) - d(c_p^{l(p)}, Z) + d(c_p^{l(p)}) - d(c_p^{l(p)-1}) + \ldots + d(\mathbf{0}, Z)$ where $l(p)$ is a cutoff level of point $p$ such that the cell containing $p$ in level $l(p)$ is heavy but no longer heavy in level $l(p) + 1$. We then sample point $p$ with some probability defined according to $l(p)$. In other words, we only sample points from heavy cells and not from non-heavy ones. Since a heavy cell contains enough points, the sampling-based estimation of the number of points is accurate enough and thus allows us to adjust them to be all positive.

Finally, to handle the insertion and deletions, we use a $F_2$-heavy hitter algorithm to identify the heavy cells. We use pseudo-random hash functions (e.g. Nisan's construction [Nis92, Ind00b] or $k$-wise independent hash functions) to do the sampling and use a `K-Set` data structure [Gan05] to store the sampled points in the dynamic stream.

## 1.3 Related Work

There is a rich history in studies of geometric problems in streaming model. Among these problems some excellent examples are: approximating the diameter of a point set [FKZ05, Ind03], approximately maitain the convex hull [CM03, HS04], the min-volume bounding box [Cha04, CS06], maintain $\epsilon$-nets and $\epsilon$-approximations of a data stream [BCEG07]. Clustering problem is another interesting and popular geometric problem studied in streaming model. There has been a lot of works on clustering data streams for the $k$-median and $k$-means problem based on coresets [HM04, HK05, Che09, FMS07, FSS13, FL11]. Additionally [CCFM97, GMMO00, Mey01] studied the problem in the more general metric space. The currently best known algorithm for $k$-median problem in this setting is an $O(1)$-approximation using $O(k\text{polylog}n)$ space [COP03]. However, all of the above methods do not work for dynamic streams.

The most relevant works to ours are those by Indyk [Ind04], Indyk & Price [IP11] and Frahling & Sohler [FS05]. Indyk [Ind04] introduced the model for dynamic geometric data streamings.

He studied algorithms for (the weight of) minimum weighted matching, minimum bichromatic matching and minimum spanning tree and $k$-median clustering. He gave a exhaustive search $(1+\epsilon)$ approximation algorithm for $k$-median and a $(\alpha, \beta)$-bi-criterion approximation algorithm. Indyk & Price [IP11] studied the problem of sparse recovery under Earth Mover Distance. They show a novel connection between EMD/EMD sparse recovery problem to $k$-median clustering problem on a two dimensional grid. The most related work to current one is Frahling & Sohler [FS05], who develop a streaming $(1+\epsilon)$-approximation algorithms for $k$-median as well as other problems over dynamic geometric data streams. All previous constructions for higher dimensional grid require space exponential in the dimension $d$.

## 2  Preliminaries

For integer $a \leq b$, we denote $[a] := \{1, 2, \ldots, a\}$ and $[a, b] := \{a, a+1, \ldots, b\}$ for integer intervals. We will consider a point set $P$ from the Euclidean space $\{1, \ldots, \Delta\}^d$. Without loss of generality, we always assume $\Delta$ is of the form $2^L$ for some integer $L$, since otherwise we can always pad $\Delta$ without loss of a factor more than 2. Our streaming algorithm will process insertions and deletions of points from this space. We study the $k$-median problem, which is to minimize $\text{cost}(P, Z) = \sum_{p \in P} d(p, Z)$ among all sets $Z$ of $k$ centers from $\mathbb{R}^d$ and where $d(p, q)$ denotes the Euclidean distance between $p$ and $q$ and $d(p, Z)$ for a set of points $Z$ denotes the distance of $p$ to the closest point in $Z$. The following definition is from [HM04].

**Definition 2.1.** *Let $P \subseteq [\Delta]^d$ be a point set. A small weighted set $S$ is called an $\epsilon$-coreset for the $k$-median problem, if for every set of $k$ centers $Z \subset [\Delta]^d$ we have* [1]

$$(1 - \epsilon) \cdot \text{cost}(P, Z) \leq \text{cost}(S, Z) \leq (1 + \epsilon) \cdot \text{cost}(P, Z),$$

*where $\text{cost}(S, Z) := \sum_{s \in S} \text{wt}(s) d(s, Z)$ and $\text{wt}(s)$ is the weight of point $s \in S$.*

Through out the paper, we assume parameters $\epsilon, \rho, \delta, \in (0, \frac{1}{2})$ unless otherwise specified. For our algorithms and constructions we define a nested grid with $L$ levels, in the following manner.

**Definition of grids** Let $v = (v_1, \ldots, v_d)$ be a vector chosen uniformly at random from $[0, \Delta - 1]^d$. Partition the space $\{1, \ldots, \Delta\}^d$ into a regular Cartesian grid $\mathcal{G}_0$ with side-length $\Delta$ and translated so that a vertex of this grid falls on $v$. Each cell of this grid can be expressed as $[v_1 + n_1 \Delta, v_1 + (n_1 + 1)\Delta) \times \ldots \times [v_d + n_d \Delta, v_d + (n_d + 1)\Delta)$ for some $(n_1, \ldots, n_d) \in \mathbb{Z}^d$. For $i \geq 1$, define the regular grid $\mathcal{G}_i$ as the grid with side-length $\Delta/2^i$ aligned such that each cell of $\mathcal{G}_{i-1}$ contains $2^d$ cells of $\mathcal{G}_i$. The finest grid is $\mathcal{G}_L$ where $L = \lceil \log_2 \Delta \rceil$; the cells of this grid therefore have side-length at most 1 and thus contain at most a single input point. Each grid forms a partition of the point-set $S$. There is a $d$-ary tree such that each vertex at depth $i$ corresponds to a cell in $\mathcal{G}_i$, and this vertex has $2^d$ children which are the cells of $\mathcal{G}_{i+1}$ that it contains. For convenience, we define $\mathcal{G}_{-1}$ as the entire dataset and it contains a single cell $\mathcal{C}_{-1}$. For each cell $\mathcal{C}$, we also treat it as a subset of the input points (i.e. $\mathcal{C} \cap P$) if there is no confusion.

We denote $Z^* \subset [\Delta]^d$ as the optimal solution for $k$-median and $\texttt{OPT}$ as the optimal cost for $Z^*$. The proof of the following lemma is delayed to Section A.

---
[1] For simplicity of the presentation, we define the coreset for all sets of k centers $Z \subset [\Delta]^d$, but it can be generalized to all sets of k centers $Z \subset \mathbb{R}^d$ with an additional polylog$(1/\epsilon)$ factor in the space. We discuss this point further in Section 6.

**Lemma 2.2.** *Fix a set $Z \subset [\Delta]^d$, then with probability at least $1 - \rho$, for every level $i \in [0, L]$, the number of cells that satisfy $d(\mathcal{C}, Z) \leq \Delta/(2^{i+1}d)$ is at most $e|Z|(L+1)/\rho$.*

## 2.1 Outline

In Section 3, we introduce the coreset with negative weights. In Section 4, we introduce a modified construction with all positive weights. Section 6 comes with the final remarks.

## 3 Generally Weighted Coreset

In this section, we present our generally weighted coreset construction. In Section 3.1, we introduce the telescope sum representation of a point $p$ and the coreset framework. In Section 3.2, we illustrate our coreset framework with an offline construction. In Section 3.3 we present an one pass streaming algorithm that implements our coreset framework.

### 3.1 The Telescope Sum and Coreset Framework

Our first technical idea is to write each point as a telescope sum. We may interpret this sum as replacing a single point by a set of points in the following way. Each term $(p - q)$ of the sum can be viewed as a pair of points $p$ and $q$, where $p$ has weight 1 and $q$ has weight $-1$. The purpose of this construction is that the contribution of each term $(p - q)$ (or the corresponding two points) is bounded. This can be later exploited when we introduce and analyze our sampling procedure.

We now start to define the telescope sum, which will relate to our nested grids. For each $\mathcal{C} \in G_i$, denote $c(\mathcal{C})$ (or simply $c$) as its center. For each point $p \in P$, define $\mathcal{C}(p, i)$ as the cell that contains $p$ in $\mathcal{G}_i$, and $c_p^i$ is the center of $\mathcal{C}(p, i)$. Then we can write

$$p = c_p^{-1} + \sum_{i=0}^{L} c_p^i - c_p^{i-1}.$$

where we set $c_p^{-1} = \mathbf{0}$ (we also call this the cell center of the $(-1)$-st level for convenience). The purpose of this can be seen when we consider the distance of $p$ to an arbitrary $k$-centers $Z \subset [\Delta]^d$, we can write the cost of a single point $p$, as

$$d(p, Z) = d(c_p^{-1}, Z) + \sum_{i=0}^{L} d(c_p^i, Z) - d(c_p^{i-1}, Z).$$

Note that $c_p^L = p$ since the cells of $\mathcal{G}_L$ contain a single point. Thus the cost of the entire set $\text{cost}(P, Z)$ can be written as,

$$\sum_{i=0}^{L} \sum_{p \in P} d(c_p^i, Z) - d(c_p^{i-1}, Z) + \sum_{p \in P} d(c_p^{-1}, Z). \tag{1}$$

As one can see, we transform the cost defined using the original set of points to the "cost" defined using cell centers. To estimate the cost, it remains to estimate each of the terms, $\sum_{p \in P} d(c_p^i, Z) - d(c_p^{i-1}, Z)$ for $i \in [0, L]$ and $\sum_{p \in P} d(c_p^{-1}, Z)$. In other words, assign weights to each of the centers of the grid cells. For $i \in [0, L]$, and a cell $\mathcal{C} \in \mathcal{G}_i$, denote $\mathcal{C}^P$ as the parent cell of $\mathcal{C}$ in grid $\mathcal{G}_{i-1}$. Thus we can rewrite the cost term as follows,

$$\text{cost}(\mathcal{G}_i, Z) := \sum_{p \in P} d(c_p^i, Z) - d(c_p^{i-1}, Z)$$

$$= \sum_{\mathcal{C} \in \mathcal{G}_i} \sum_{p \in \mathcal{C}} d(c(\mathcal{C}), Z) - d(c(\mathcal{C}^P), Z)$$

$$= \sum_{\mathcal{C} \in \mathcal{G}_i} |\mathcal{C}| \left[ d(c(\mathcal{C}), Z) - d(c(\mathcal{C}^P), Z) \right]$$

$$= \sum_{\mathcal{C} \in \mathcal{G}_i} |\mathcal{C}| d(c(\mathcal{C}), Z) - \sum_{\mathcal{C}' \in \mathcal{G}_{i-1}} \sum_{\mathcal{C} \in \mathcal{G}_i : \mathcal{C} \subset \mathcal{C}'} |\mathcal{C}| d(c(\mathcal{C}'), Z). \tag{2}$$

For $i = -1$, we denote $\text{cost}(\mathcal{G}_{-1}, Z) = |P| d(c_p^{-1}, Z)$. Then this leads to our following coreset construction framework.

**Generally Weighted Construction** The coreset $S$ in the construction is composed by a weighted subset of centers of grid cells. The procedure of the construction is to assign some (integer) value to each cell center. For instance, maintain a integer valued function $\widehat{|\cdot|}$ on cells (using small amount of space). $\widehat{|\mathcal{C}|}$ is called the *value* of the cell $\mathcal{C}$. Let $c$ be the center of $\mathcal{C}$, then the weight for $c$ is

$$\text{wt}(c) = \widehat{|\mathcal{C}|} - \sum_{\mathcal{C}' : \mathcal{C}' \in \mathcal{G}_{i+1}, \mathcal{C}' \subset \mathcal{C}} \widehat{|\mathcal{C}'|}. \tag{3}$$

And for the $L$-th grid $\mathcal{G}_L$, the weight for each cell $\mathcal{C}$ is just $\widehat{|\mathcal{C}|}$. Note that there might be negative weights for some cells.

As a naïve example, we set $\widehat{|\mathcal{C}|} := |\mathcal{C}|$ as the exact number of points of a cell $\mathcal{C}$. Then we would expect the cells in every level except those in $\mathcal{G}_L$ have weight 0. In other words, we stored the entire point set as the coreset. As we will show, if we allow $\widehat{|C|}$ as an approximation of $|C|$ up to additive error, we can compress the number of non-zero weighted centers to be a smaller number.

**Definition 3.1.** *Given a grid structure, and a real valued function $\widehat{|\cdot|}$ on the set of cells. We define a function $\widehat{\text{cost}} : [\Delta]^d \times \mathcal{G} \to \mathbb{R}$ as follows, for $i \in [0, L]$ and $Z \subset [\Delta]^d$,*

$$\widehat{\text{cost}}(\mathcal{G}_i, Z) := \sum_{\mathcal{C} \in \mathcal{G}_i} \widehat{|\mathcal{C}|} d(c(\mathcal{C}), Z) - \sum_{\mathcal{C}' \in \mathcal{G}_{i-1}} \sum_{\mathcal{C} \in \mathcal{G}_i : \mathcal{C} \subset \mathcal{C}'} \widehat{|\mathcal{C}|} \cdot d(c(\mathcal{C}'), Z), \tag{4}$$

*and $\widehat{\text{cost}}(\mathcal{G}_{-1}, Z) = \widehat{|\mathcal{C}_{-1}|} d(\mathbf{0}, Z)$, where $\mathcal{C}_{-1}$ is the cell in $\mathcal{G}_{-1}$ containing the entire set of points.*

**Lemma 3.2.** *Fix an integer valued function $\widehat{|\cdot|}$ on the set of cells and parameter $0 < \epsilon < \frac{1}{2}$. Let $S$ be the set of all cell centers with weights assigned by Equation (3). If $\widehat{|\mathcal{C}_{-1}|} = |P|$ (recall that $\mathcal{C}_{-1}$ is the first cell containing the entire dataset) and for any $Z \subset [\Delta]^d$ with $|Z| \leq k$ and $i \in [0, L]$*

$$\left| \text{cost}(\mathcal{G}_i, Z) - \widehat{\text{cost}}(\mathcal{G}_i, Z) \right| \leq \frac{\epsilon \text{OPT}}{L+1},$$

*then $S$ is an $\epsilon$-coreset for $k$-median.*

*Proof.* Given an arbitrary set of centers $Z \subset [\Delta]^d$,

$$\begin{aligned}
\text{cost}(S, Z) &= \sum_{s \in S} \text{wt}(s) d(s, Z) \\
&= \sum_{i \in [0,L]} \sum_{\mathcal{C} \in \mathcal{G}_i} d(c(\mathcal{C}), Z) \left( \widehat{|\mathcal{C}|} - \sum_{\substack{\mathcal{C}': \mathcal{C}' \in \mathcal{G}_{i+1} \\ \mathcal{C}' \subset \mathcal{C}}} \widehat{|\mathcal{C}'|} \right) + |P| d(\mathbf{0}, Z) \\
&= \sum_{i \in [0,L]} \Big[ \sum_{\mathcal{C} \in \mathcal{G}_i} \widehat{|\mathcal{C}|} d(c(\mathcal{C}), Z) - \sum_{\substack{\mathcal{C}' \in \mathcal{G}_{i-1} \\ \mathcal{C} \in \mathcal{G}_i : \mathcal{C} \subset \mathcal{C}'}} \widehat{|\mathcal{C}|} d(c(\mathcal{C}'), Z) \Big] + |P| d(\mathbf{0}, Z) \\
&= \sum_{i \in [0,L]} \widehat{\text{cost}}(\mathcal{G}_i, Z).
\end{aligned}$$

It follows that $|\text{cost}(S, Z) - \text{cost}(P, Z)| \leq \epsilon \mathsf{OPT}$. □

## 3.2 An Offline Construction

In this section, we assume we have $(10, 10)$-bi-criterion approximation to $k$-median. Let $Z' = \{z'_1, z'_2, \ldots, z'_{10k}\}$ be the centers and $o$ is the cost satisfying $\mathsf{OPT} \leq o \leq 10\mathsf{OPT}$. This can be done using [Ind00a]. We will show how we construct the coreset base on the framework described in the last section.

**An Offline Construction** For each point in level $\mathcal{G}_{-1}$, we sample it with probability $\pi_{-1} = 1$ (i.e. count the number of points exactly) and set $\widehat{|\mathcal{C}_{-1}|} := |P|$. For each level $i \in [0, L]$, we pick the set of all cells $\mathcal{C}$ satisfying $d(\mathcal{C}, Z') \leq W/(2d)$, where $W$ is the side length of $\mathcal{C}$. Denote the set of these cells as $C_{Z'}$. We count the number of points in each of these cells exactly, and set $\widehat{|\mathcal{C}|} := |\mathcal{C}|$. For the points in the rest of cells, for each $i \in [0, L]$, we sample the points with probability

$$\pi_i = \min\left( \frac{200(L+1)^2 \Delta d^2}{2^i \epsilon^2 o} \ln \frac{2(L+1) \Delta^{kd}}{\rho}, 1 \right) \tag{5}$$

uniformly and independently. Denote $S_i$ as the set of sampled points at level $i$. For each $\mathcal{C} \notin C_{Z'}$, set

$$\widehat{|\mathcal{C}|} := |S_i \cap \mathcal{C}| / \pi_i.$$

Then, from the bottom level to the top level, we assign the weight to the cell centers of each of the cells and their parent cells with non-zero $\widehat{|\mathcal{C}|}$ using (3). Denote $\mathcal{S}$ as the coreset, which contains the set of cell centers of non-zero weight.

**Theorem 3.3.** *Fix $\epsilon, \rho \in (0, 1/2)$, then with probability at least $1 - 8\rho$, the offline construction $\mathcal{S}$ is an $\epsilon$-coreset for $k$-median and that*

$$|\mathcal{S}| = O\left( \frac{d^4 k L^4}{\epsilon^2} \log \frac{1}{\rho} + \frac{kL^2}{\rho} \right).$$

*Proof of Theorem 3.3.* By definition $\widehat{|\mathcal{C}_{-1}|} = |P|$, it is suffice to show that with probability at least $1 - 4\rho$, for every $i \in [0, L]$ and every $k$-set $Z \subset [\Delta]^d$,

$$\left| \widehat{\text{cost}}(\mathcal{G}_i, Z) - \text{cost}(\mathcal{G}_i, Z) \right| \leq \epsilon \mathsf{OPT}/(L+1).$$

It follows from Lemma 3.2 that, $\mathcal{S}$ is an $\epsilon$-coreset.

Let $S_i$ be the sampled points of level $i$. Fix a $k$-set $Z \subset [\Delta]^d$, for each $i \in [0, L]$, by equation (2), we have that, $\widehat{\text{cost}}(\mathcal{G}_i, Z) = \sum_{\mathcal{C} \in C_{Z'}} |\widehat{\mathcal{C}}| \left(d(c(\mathcal{C}), Z) - d(c(\mathcal{C}^P), Z)\right) + \sum_{p \in S_i} (d(c_p^i, Z) - d(c_p^{i-1}, Z)))/\pi_i$.
Note that $E(\widehat{\text{cost}}(\mathcal{G}_i, Z)) = \text{cost}(\mathcal{G}_i, Z)$. The first term contributes 0 to the difference $\widehat{\text{cost}}(\mathcal{G}_i, Z) - \text{cost}(\mathcal{G}_i, Z)$ since each $|\widehat{\mathcal{C}}|$ is exact. It remains to bound the error contribution from the second part. Denote $A_2 = \sum_{p \in S_i} (d(c_p^i, Z) - d(c_p^{i-1}, Z)))/\pi_i$. Recall that $Z'$ is the centers of the bi-criterion solution and $C_{Z'}$ is the set of cells with distance less than $W/(2d)$ to $Z'$, where $W$ is the side-length of a cell. Let $\mathcal{A}$ be event that $|C_{Z'}| \leq e|Z'|(L+1)^2/\rho = O(kL^2/\rho)$. By Lemma 2.2, $\mathcal{A}$ happens with probability at least $1 - \rho$. Conditioning on $\mathcal{A}$ happening, for each point $p \in \mathcal{C} \notin C_{Z'}$, we have that $d(p, Z') \geq \text{diam}(\mathcal{C})/(2d^{3/2})$. Therefore, $\sum_{p \in \mathcal{C} \notin C_Z} \text{diam}(\mathcal{C}) \leq (2d^{3/2}) \sum_{p \in \mathcal{C} \notin C_Z} d(p, Z') \leq 20d^{3/2}\text{OPT}$. By Lemma 3.4, with probability at least $1 - \frac{\rho}{(L+1)\Delta^{kd}}$, $|A_2 - E(A_2)| \leq \frac{\epsilon \text{OPT}}{L+1}$. Since there are at most $\Delta^{kd}$ many different $k$-sets from $[\Delta]^d$, thus, for a fixed $i \in [0, L]$ with probability at least $1 - \frac{\rho}{L+1}$, for all $k$-sets $Z \subset [\Delta]^d$, $\left|\widehat{\text{cost}}(\mathcal{G}_i, Z) - \text{cost}(\mathcal{G}_i, Z)\right| \leq \epsilon\text{OPT}/(L+1)$. By the union bound, with probability at least $1 - 4\rho$, $\mathcal{S}$ is the desired coreset.

It remains to bound the size of $\mathcal{S}$. Conditioning on $\mathcal{A}$ happening, then $|C_{Z'}| = O(kL^2/\rho)$. For each level $i$, since each point from cells $\mathcal{C} \notin C_{Z'}$ contributes at least $\Delta/(2^{i+1}d)$ to the bi-criterion solution, there are at most $O(2^i\text{OPT}d/\Delta)$ points in cells not in $C_{Z'}$. By a Chernoff bound, with probability at least $1 - \rho/(L+1)$, the number of points sampled from cells $\mathcal{C} \notin C_{Z'}$ of level $i$ is upper bounded by $O(d^4kL^3 \log \frac{1}{\rho}/\epsilon^2)$. Thus for all levels, with probability at least $1 - \rho$, the number of points sampled is upper bounded by $O(d^4kL^4 \log \frac{1}{\rho}/\epsilon^2)$, which is also an upper bound of the number of cells occupied by sampled points. Now we bound the number of non-zero weighted centers. In the coreset construction, if a cell center has non-zero weight, then either itself or one of its children cells has non-zero assigned value $|\widehat{\mathcal{C}}|$. Thus the number of non-zero weigted centers is upper bound by 2 times the number of non-zero valued cells. Thus $|\mathcal{S}| = O(d^4kL^4 \log \frac{1}{\rho}/\epsilon^2 + \frac{kL^2}{\rho})$. $\square$

**Lemma 3.4.** *Fix $\epsilon, \rho \in (0, 1/2)$, if a set of cells $C$ from grid $\mathcal{G}_i$ satisfies $\sum_{\mathcal{C} \in C} |\mathcal{C}|\text{diam}(\mathcal{C}) \leq \beta\text{OPT}$ for some $\beta \geq 2\epsilon/(3(L+1))$, let $S$ be a set of independent samples from the point set $\cup \{\mathcal{C} \in C\}$ with probability*
$$\pi_i \geq \min\left(\frac{3a(L+1)^2 \Delta \sqrt{d}\beta}{2^i \epsilon^2 o} \ln \frac{2\Delta^{kd}(L+1)}{\rho}, 1\right)$$
*where $0 < o \leq a\text{OPT}$ for some $a > 0$, then for a fixed set $Z \subset [\Delta]^d$, with probability at least $1 - \rho/((L+1)\Delta^{kd})$,*
$$\left|\sum_{p \in S}(d(c_p^i, Z) - d(c_p^{i-1}, Z))/\pi_i - \sum_{p \in \cup\{\mathcal{C} \in C\}}(d(c_p^i, Z) - d(c_p^{i-1}, Z))\right| \leq \frac{\epsilon\text{OPT}}{L+1}.$$

The proof is a straightforward application of Bernstein inequality. It is presented in Section A.

### 3.3 The Streaming Algorithm

For the streaming algorithm, the first challenge is that we do not know the actual value of OPT, neither do we have an $(\alpha, \beta)$-bi-criterion solution. To handle this, we will show that we do not need an actual set of centers of an approximate solution, and that a conceptual optimal solution suffices. We will guess logarithmically many values for OPT to do the sampling. We re-run the algorithm in parallel for each guess of OPT.

The second challenge is that we cannot guarantee the sum $\sum_{\mathcal{C} \in \mathcal{G}_i} \mathrm{diam}(\mathcal{C})$ to be upper bounded by $\beta\mathsf{OPT}$ as required in Lemma 3.4. We will show that we can split the set of cells into two parts. The first part satisfies the property that $\sum_{\mathcal{C}} |\mathcal{C}|\mathrm{diam}(\mathcal{C}) \leq \beta\mathsf{OPT}$ for some parameter $\beta$. The second part satisfies that $|\mathcal{C}|\mathrm{diam}(\mathcal{C}) \geq a\mathsf{OPT}/k$ for some constant $a$.

For the first part, we use a similar sampling procedure as we did in the offline case. The challenge here is that there might be too many points sampled when the algorithm is midway through the stream, and these points may be deleted later in the stream. To handle this case, we use a data structure called K-Set structure with parameter $k$ [Gan05]. We will insert (with deletions) a multiset of points $M \subset [N]$ into the K-Set. The data structure processes each stream operation in $O(\log(k/\delta))$ time. At each point of time, it supports an operation RETRSET, that with probability at least $1-\delta$ either returns the set of items of $M$ or returns Fail. Further, if the number of distinct items $|M|$ is at most $k$, then RETRSET returns $M$ with probability at least $1 - \delta$. The space used by the K-Set data structure is $O(k(\log |M| + \log N) \log(k/\delta))$. The K-Set construction also returns the frequency of each stored points upon the RETRSET operation.

For the second part, we call these cells *heavy*. We first upper bound the number of heavy cells by $\alpha k$ for some $\alpha > 1$. We use a heavy hitter algorithm HEAVY-HITTER to retrieve an approximation to the number of points in these cells. The guarantee is given in the following theorem. In an insertion-deletion stream, it may that although the stream has arbitrary large length, at any moment a much smaller number of elements are active (that is, inserted and not yet deleted). We define the size of a stream to be the maximum number of active elements at any point of the stream.

**Theorem 3.5** ([LNNT16] Theorem 2). *Fix $\epsilon, \delta \in (0, 1/2)$. Given a stream (of insertions and deletions) of size $m$ consisting of items from universe $[n]$, there exists an algorithm HEAVY-HITTER$(n, k, \epsilon, \delta)$ that makes a single pass over the stream and outputs a set of pairs $H$. With probability at least $1 - \delta$, the following holds,*
*(1) for each $(i, \hat{f}_i) \in H$, $f_i^2 \geq \sum_{j=1}^n f_j^2/k - \epsilon^2 \sum_{j=k+1}^n f_j^2$;*
*(2) if for any $i \in [n]$ and $f_i^2 \geq \sum_{j=1}^n f_j^2/k + \epsilon^2 \sum_{j=k+1}^n f_j^2$, then $(i, \hat{f}_i) \in H$;*
*(3) for each $(i, \hat{f}_i) \in H$, $|\hat{f}_i - f_i| \leq \epsilon \sqrt{\sum_{j=k+1}^n f_j^2}$.*
*The algorithm uses $O\left((k + \frac{1}{\epsilon^2}) \log \frac{n}{\delta} \log m\right)$ bits of space, $O(\log n)$ update time and $O(k+1/\epsilon^2)\mathrm{polylog}(n)$ query time.*

Thus, using HEAVY-HITTER, we are guaranteed that the error of the number of points in heavy cells is upper bounded by $\epsilon$ times the number of points in the non-heavy cells. The first heavy hitter algorithm that achieves an $l_2$ guarantee is by [CCFC02], who has the same space and update time as that of the above algorithm. However the update time is slow, i.e. $O(n \log n)$ time to output the set of heavy hitters.

Lastly, we will use fully independent random hash function to sample the points. We will use Nissan's pseudorandom generator to de-randomize the hash functions by the method of [Ind00b]. Our main theorem for this section is as follows. The formal proof of this theorem is postponed to Section B.

**Theorem 3.6** (Main Theorem). *Fix $\epsilon, \rho \in (0, 1/2)$, positive integers $k$ and $\Delta$, Algorithm 1 makes a single pass over the streaming point set $P \subset [\Delta]^d$, outputs a weighted set $S$, such that with probability at least $1 - \rho$, $S$ is an $\epsilon$-coreset for k-median of size $O\left(\frac{d^4 L^4 k}{\epsilon^2} + \frac{L^2 k}{\rho}\right)$, where $L = \log \Delta$. The algorithm uses*
$$O\left[k\left(\left(\frac{d^7 L^7}{\epsilon^2} + \frac{d^3 L^5}{\rho}\right) \log \frac{dkL}{\rho\epsilon} + \frac{d^5 L^6}{\epsilon^2 \rho}\right)\right]$$

bits in the worst case, processes each update in time $O\left(dL^2 \log \frac{dkL}{\rho\epsilon}\right)$ and outputs the coreset in time $\text{poly}(d, k, L, 1/\epsilon)$ after one pass of the stream.

## 4 Positively Weighted Coreset

In this section, we will introduce a modification to our previous coreset construction, which leads to a coreset with all positively weighted points. The full algorithm and proofs are postponed to Section C. We present the main steps in this section.

The high level idea is as follows. When considering the estimate of the number of points in a cell, the estimate is only accurate when it truly contains a large number of points. However, in the construction of the previous section, we sample from each cell of each level, even though some of the cells contain a single point. For those cells, we cannot adjust their weights from negative to positive, since doing so would introduce large error. In this section, we introduce an ending level to each point. In other words, the number of points of a cell is estimated by sampling only if it contains many points. Thus, the estimates will be accurate enough and allow us to rectify the weights to be all positive.

### 4.1 Reformulation of the Telescope Sum

**Definition 4.1.** *A* heavy cell identification scheme *$\mathcal{H}$ is a map $\mathcal{H}: \mathcal{G} \to \{\text{heavy, non-heavy}\}$ such that, $h(\mathcal{C}_{-1}) = $ heavy and for cell $\mathcal{C} \in \mathcal{G}_i$ for $i \in [0, L]$*
1. *if $|\mathcal{C}| \geq \frac{2^i \rho d \mathsf{OPT}}{k(L+1)\Delta}$ then $\mathcal{H}(\mathcal{C}) = $ heavy;*
2. *If $\mathcal{H}(\mathcal{C}) = $ non-heavy, then $\mathcal{H}(\mathcal{C}') = $ non-heavy for every subsell $\mathcal{C}'$ of $\mathcal{C}$.*
3. *For every cell $\mathcal{C}$ in level $L$, $\mathcal{H}(\mathcal{C}) = $ non-heavy.*
4. *For each $i \in [0, L]$, $|\{\mathcal{C} \in \mathcal{G}_i : \mathcal{H}(\mathcal{C}) = \text{heavy}\}| \leq \frac{\lambda_1 kL}{\rho}$, where $\lambda_1 \leq 10$ is a positive universal constant.*

*The output for a cell not specified by the above conditions can be arbitrary. We call a cell* heavy *if it is identified heavy by $\mathcal{H}$. Note that a heavy cell does not necessarily contain a large number of points, but the total number of these cells is always bounded.*

In the sequel, heavy cells are defined by an arbitrary fixed identification scheme unless otherwise specified.

**Definition 4.2.** *Fix a heavy cell identification scheme $\mathcal{H}$. For level $i \in [-1, L]$, let $\mathcal{C}(p, i) \in \mathcal{G}_i$ be the cell in $\mathcal{G}_i$ containing $p$. The* ending level *$l(p)$ of a point $p \in P$ is the largest level $i$ such that $\mathcal{H}(\mathcal{C}(p, i)) = $ heavy, and $\mathcal{H}(\mathcal{C}(p, i+1)) = $ non-heavy.*

Note that the ending level is uniquely defined if a heavy cell identification scheme is fixed. We now rewrite the telescope sum for $p$ as follows,

$$p = \sum_{i=0}^{l(p)} \left(c_p^i - c_p^{i-1}\right) + c_p^L - c_p^{l(p)},$$

where $c_p^{-1} = \mathbf{0}$ and $c_p^L = p$. For arbitrary $k$-centers $Z \subset [\Delta]^d$, we write, $d(p, Z) = \sum_{i=0}^{l(p)} \left(d(c_p^i, Z) - d(c_p^{i-1}, Z)\right) + d(c_p^L, Z) - d(c_p^{l(p)}, Z) + d(\mathbf{0}, Z)$ .

## 4.2 The New Construction (with arbitrary weights)

For these heavy cells, we use `HEAVY-HITTER` algorithms to obtain accurate estimates of the number of points in these cells, thus providing a *heavy cell identification scheme*. For the non-heavy cells, we only need to sample points from the bottom level, $\mathcal{G}_L$, but with a different probability for points with different ending levels.

We now describe the new construction. This essentially has the same gaurantee as the simpler construction from the previous section, however the benefit here is that (as shown in the next subsection) it can be modified to output only positive weights. In the following paragraph, the estimations $\widehat{|\mathcal{C}|}$ are given as a blackbox. In proposition C.9 we specify the conditions these estimations must satisfy.

**Non-Negatively Weighted Construction** Fix an arbitrary heavy cell identification scheme $\mathcal{H}$. Let $P_l$ be all the points with ending level $l(p) = l$. For each heavy cell $\mathcal{C}$, let $\widehat{|\mathcal{C}|}$ be an estimation of number of points of $|\mathcal{C}|$, we also call $\widehat{|\mathcal{C}|}$ the *value* of cell $\mathcal{C}$. For each non-heavy cell $\mathcal{C}'$, let $\widehat{|\mathcal{C}'|} = 0$. Let $S$ be a set samples of $P$ constructed as follows: $S = S_{-1} \cup S_0 \cup S_1, \cup \ldots \cup S_L$, where $S_l$ is a set of i.i.d samples from $P_l$ with probability $\pi_l$. Here $\pi_l$ for $l \in [-1, L]$ is redefined as $\pi_l =$

$$\min\left(\frac{\lambda_3 d^2 \Delta L^2}{2^l \epsilon^2 o} \log\left(\frac{2L\Delta^{dk}}{\rho}\right) + \frac{\lambda_4 d^2 k L^3 \Delta}{2^i \epsilon^2 \rho o} \log \frac{30kL^2}{\rho^2}, 1\right)$$

where $\lambda_3 > 0$ and $\lambda_4 > 0$ are universal constants. Our coreset $\mathcal{S}$ is composed by all the sampled points in $S$ and the cell centers of heavy cells, with each point $p$ assigned a weight $1/\pi_{l(p)}$ and for each cell center $c$ of a heavy cell $\mathcal{C} \in \mathcal{G}_i$, the weight is,

$$\text{wt}(c) = \widehat{|\mathcal{C}|} - \sum_{\substack{\mathcal{C}': \mathcal{C}' \in \mathcal{G}_{i+1}, \mathcal{C}' \subset \mathcal{C}, \\ \mathcal{C}' \text{ is heavy}}} \widehat{|\mathcal{C}'|} - \frac{|S_i \cap \mathcal{C}|}{\pi_i}. \tag{6}$$

For each non-heavy cell $\mathcal{C}$ except for those in the bottom level, $\text{wt}(c(\mathcal{C})) = 0$. The weight of each point from $S$ is the value of the corresponding cell in the bottom level.

## 4.3 Ensuring Non-Negative Weights

We now provide a procedure to rectify all the weights for the coreset constructed in the last subsection. The idea is similar to the method used in [IP11]. The procedure is shown in Algorithm 4.3. After this procedure, there will be no negative weights in the coreset outputs.

**Theorem 4.3.** *Fix $\epsilon, \rho \in (0, 1/2)$, positive integers $k$ and $\Delta$, Algorithm 6 makes a single pass over the streaming point set $P \subset [\Delta]^d$, outputs a weighted set $S$ with non-negative weights for each point, such that with probability at least $0.99$, $S$ is an $\epsilon$-coreset for $k$-median of size*

$$O\left[\frac{d^3 L^4 k}{\epsilon^2}\left(d + \frac{1}{\rho}\log\frac{kL}{\rho}\right)\right]$$

*where $L = \log \Delta$. The algorithm uses*

$$O\left[\frac{d^7 L^7 k}{\epsilon^2}\left(\rho dL + \frac{L}{\rho}\log^2 \frac{dkL}{\rho\epsilon}\right)\log^2 \frac{dkL}{\rho\epsilon}\right]$$

*bits in the worst case. For each update of the input, the algorithm needs $\text{poly}(d, 1/\epsilon, L, \log k)$ time to process and outputs the coreset in time $\text{poly}(d, k, L, 1/\epsilon, 1/\rho, \log k)$ after one pass of the stream.*

# 5 Experiments

We illustrate our construction using an offline construction on Gaussian mixture data in $\mathbb{R}^2$. As shown in Figure 2 in Section D, we randomly generated 65536 points from $\mathbb{R}^2$, then rounded the points to a grid of size $\Delta = 512$. Our coreset uses $\log_2 \Delta + 2 = 11$ levels of grids. The storage in each level is very sparse. As shown in Figure 1(a), only 90 points are stored in total. We compared the 1-median costs estimated using the coreset and the dataset, the resulting difference is very small, as illustrated in Figure 1(b).

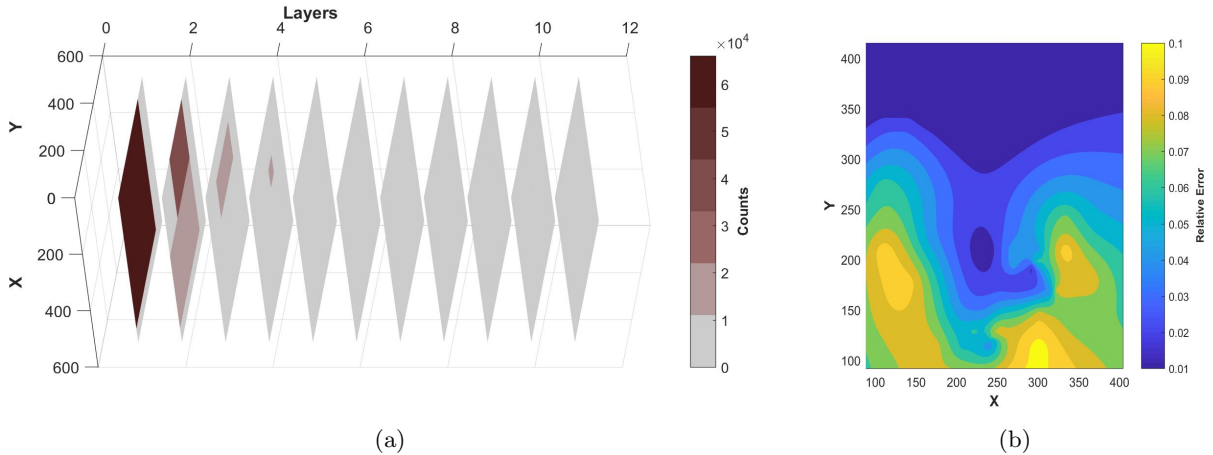

(a) (b)

Figure 1: (a) The layer structure of the coreset. Cells with more weight are shaded darker. (b) The relative error of a 1-median cost function. Using only 90 points, the global maximum error was under 10%.

# 6 Concluding Remark

We develop algorithms that make a single pass over the dynamic stream and output, with high probability, a coreset for the original $k$-median problem. Both the space complexity and the size of the coreset are polynomially dependent on $d$, whereas the only previous known bounds are exponential in $d$. We constructed our coreset for the possible solutions in discrete space $[\Delta]^d$, but it is easy to modify the coreset to be a coreset in continuous space $[0, \Delta]^d$ (note that we still require the input dataset to be from a discrete space). The way to do this is by modifying the sampling probability $\pi_i$ in the algorithm, i.e. replacing the factor of $\ln(\Omega(\Delta^{kd} L/\rho))$ to $\ln(\Omega((\Delta/\epsilon)^{kd} L/\rho))$. Then any $k$-set from $[0, \Delta]^d$ can be rounded to the closest $k$-set in $[\Delta/\epsilon]^d$ and the cost only differs by a $(1 \pm \epsilon)$ factor while the space bound changes only by a polylog$(1/\epsilon)$ factor. Lastly, we remark that the coreset scheme can be easily modified to other metric spaces, e.g. the $l_p$ metric. The space bound depends on the doubling dimension of the metric.

As shown in our experiments, a 2D implementation using our framework is very efficient. We believe that a high-dimensional implementation will be efficient as well. We leave the full implementation as a future project.

**Algorithm 1** CoreSet$(S, k, \rho, \epsilon)$: construct a $\epsilon$-coreset for dynamic stream $S$.

**Initization**:
Initialize a grid structure;
$O \leftarrow \{1, 2, 4, \ldots, \sqrt{d}\Delta^{d+1}\}$;
$L \leftarrow \lceil \log \Delta \rceil$;
$\pi_i(o) \leftarrow \min\left(\frac{3(L+1)^2 \Delta d^2}{2^i \epsilon^2 o} \ln \frac{2\Delta^{kd}(L+1)}{\rho}, 1\right)$;
$K \leftarrow \frac{(2+e)(L+1)k}{\rho} + \frac{24d^4(L+1)^3 k}{\epsilon^2} \ln \frac{1}{\rho}$,
$\epsilon' \leftarrow \left(\epsilon \sqrt{\frac{\rho}{8(2+e)^2 k d^3 (L+1)^3}}\right)$; $m \leftarrow 0$;
For each $o \in O$ and $i \in [0, L]$, construct fully independent hash function $h_{o,i} : [\Delta]^d \to \{0, 1\}$ with $Pr_{h_{o,i}}(h_{o,i}[q] = 1) = \pi_i(o)$;
Initialize K-Set instances $\text{KS}_{o,i}$ with error probability $\rho/(L+1)$, size parameter $K$;
Initialize HEAVY-HITTER$(\Delta^d, (e+2)(L+1)k/\rho, \epsilon', \rho/(L+1))$ instances, $\text{HH}_0, \text{HH}_1, \ldots, \text{HH}_L$, one for a level;

**Update** $(S)$:
**for** *each update* $(op, q) \in S$:
    /\*$op \in \{\text{Insert}, \text{Delete}\}$\*/
    $m \leftarrow m \pm 1$; /\*Insert: $+1$, Delete:$-1$\*/
    **for** *each* $i \in [0, L]$:
        $c_q^i \leftarrow$ the center of the cell contains $q$ at level $i$;
        $\text{HH}_i.\text{update}(op, c_q^i)$;
        **for** *each* $o \in [O]$:
            **if** $h_{o,i}(q) == 1$:
                $\text{KS}_{o,i}.\text{update}(op, c_q^i)$;

**Query**:
Let $o^*$ be the smallest $o$ such that no instance of $\text{KS}_{o,0}, \text{KS}_{o,1}, \ldots, \text{KS}_{o,L}$ returns Fail;
$R \leftarrow \{\}$;
**for** $i = -1$ *to* $L$:
    **for** *each cell center $c$ in level $i$*:
        Let $\mathcal{C}$ be the cell containing $c$;
        **if** $i = -1$:
            $f \leftarrow m$;
        **else**:
            $f \leftarrow \text{GetFreq}(c, \text{HH}_i, \text{KS}_{o^*,i}, \pi_i(o^*))$;
        **if** $i < L$:
            $g \leftarrow \sum_{\mathcal{C}' \subset \mathcal{C}: \mathcal{C}' \in \mathcal{G}^{i+1}}$
                $\text{GetFreq}(c(\mathcal{C}'), \text{HH}_{i+1}, \text{KS}_{o^*i+1}, \pi_i(o^*))$;
            Assign weight $f - g$ to $c$;
            **if** $f - g \neq 0$:
                $R \leftarrow R \cup \{c\}$;
        **else**:
            Assign weight $f$ to $c$;
            **if** $f \neq 0$:
                $R \leftarrow R \cup \{c\}$;
**return** $R$.

**Algorithm 2** GetFreq($e, \text{HH}, \text{KS}, \pi_i$): retrieve the correct frequency of cell center $e$, given the instance of HEAVY-HITTER and K-set.

---

$f_S(e) \leftarrow$ the frequency of $e$ returned by HH;
$f_K(e) \leftarrow$ the frequency of $e$ returned by KS;
$k' \leftarrow (e+2)(L+1)k/\rho$;
$F \leftarrow$ the set of top-$k'$ heavy hitters returned by HEAVY-HITTER;
**if** $e \in F$**:**
$\quad$ **return** $f_S(e)$;
**else:**
$\quad$ **return** $f_K(e)/\pi_i$.

---

**Algorithm 3** RectifyWeights$\left(\widehat{|\mathcal{C}_1|}, \widehat{|\mathcal{C}_2|} \ldots, \widehat{|\mathcal{C}_{k'}|}, S\right)$: input the estimates of number of points in each cell and the weighted sampled points, output a weighted coreset with non-negative weights.

---

**for** $i = -1$ *to* $L$**:**
$\quad$ **for** *each heavy cell $\mathcal{C}$ center in $\mathcal{G}_i$*:
$\quad\quad$ **if** $\text{wt}(\mathcal{C}) < 0$**:**
$\quad\quad\quad$ Decrease the value of the children heavy cells in level $\mathcal{G}_{i+1}$ and sampled points $S_i$ arbitrarily by total $|\text{wt}(\mathcal{C})|$ amount, such that for each children cell $\mathcal{C}' \in \mathcal{G}_{i+1}$, $\widehat{|\mathcal{C}'|}$ is non-negative, and for each sampled point $p \in S_i$, the weight is non-negative.
**return** *Rectified Coreset*

# A  Proofs of Section 3

*Proof of Lemma 2.2.* Fix an $i$ and consider a grid $\mathcal{G}_i$. For each center $z_j$, denote $X_{j,\alpha}$ the indicator random variable for the event that the distance to the boundary in dimension $\alpha$ of grid $\mathcal{G}_i$ is at most $\Delta/(2^{i+1}d)$. Since in each dimension, if the center is close to a boundary, it contributes a factor at most 2 to the total number of close cells. It follows that the number of cells that have distance at most $\Delta/(2^{i+1}d)$ to $z_j$ is at most,

$$N = 2^{\sum_{\alpha=1}^{d} X_{j,\alpha}}.$$

Defining $Y_{j,\alpha} = 2^{X_{j,\alpha}}$, we obtain,

$$E[N] = E\left[\prod_{\alpha=1}^{d} Y_{j,\alpha}\right] = \prod_{\alpha=1}^{d} E[Y_{j,\alpha}].$$

We have that $Pr[X_{j,\alpha} = 1] \leq 1/d$ and so we get,

$$E[Y_{j,\alpha}] \leq E[1 + X_{j,\alpha}] = 1 + E[X_{j,\alpha}] \leq 1 + 1/d.$$

Thus $E(N) = \prod_{\alpha=1}^{d} E[Y_{j,\alpha}] \leq (1+1/d)^d \leq e$. Thus the expected number of center cells is at most $(1+1/d)^d |Z| \leq e|Z|$. By Markov's inequality, the probability that we have more than $e|Z|(L+1)/\rho$ center cells in each grid is at most $\rho/(L+1)$. By a union bound, the probability that in any grid we have more than $e|Z|(L+1)/\rho$ center cells is at most $\rho$. □

*Proof of Lemma 3.4.* Let $L' = L + 1$. Note that for each point $p \in P$, $|d(c_p^i, Z) - d(c_p^{i-1}, Z)| \leq \Delta\sqrt{d}/2^i$. Denote $\hat{A} = \sum_{p \in S}(d(c_p^i, Z) - d(c_p^{i+1}, Z))/\pi_i$ and $A = \sum_{p \in \cup\{\mathcal{C} \in C\}}(d(c_p^i, Z) - d(c_p^{i+1}, Z))$. We have that $E(\hat{A}) = A$. Let

$$X_p := \mathbb{I}_{p \in S}(d(c_p^i, Z) - d(c_p^{i+1}, Z))/\pi_i,$$

where $\mathbb{I}_{p \in S}$ is the indicator function that $p \in S$. Then we have that $Var(X_p) \leq \Delta^2 d/(4^i \pi_i)$ and $b := \max_p |X_p| \leq \Delta\sqrt{d}/(2^i \pi_i)$. By Bernstein's inequality,

$$Pr\left[|\hat{A} - A| > t\right] \leq 2e^{-\frac{t^2}{2|P|\Delta^2 d/(4^i \pi_i) + 2bt/3}}$$
$$\leq 2e^{-\frac{3 \times 2^{i-1} t^2 \pi_i}{(\beta\mathsf{OPT}+\frac{t}{3})\Delta\sqrt{d}}}. \tag{7}$$

By setting $t = \epsilon\mathsf{OPT}/L'$, we have that

$$Pr\left[|\hat{A} - A| > \frac{\epsilon\mathsf{OPT}}{L'}\right] \leq 2e^{-\ln\frac{2L'\Delta^{dk}}{\rho}} \leq \frac{\rho}{L'\Delta^{dk}}. \tag{8}$$

Thus with probability $1-\rho/(L'\Delta^{dk})$, $\hat{A}$ is an $\epsilon\mathsf{OPT}/L'$ additive approximation to the sum $\sum_{p \in P} d(c_p^i, Z) - d(c_p^{i+1}, Z)$. □

# B Proof of Theorem 3.6

Before we prove this theorem, we first present Lemma B.1 and Lemma B.2. In Algorithm 1, for each level $i \in [0, L]$, let $H_i$ be the set of cells in $\mathcal{G}_i$ whose frequencies are returned by HEAVY-HITTER in the RetrieveFrequency procedure. For each $\mathcal{C} \in H_i$, let $\widehat{|\mathcal{C}|}$ be the returned frequency of $\mathcal{C}$. Let $H'_i$ be the set of cells in $\mathcal{G}_i$ whose frequencies are returned by a K-set in the RetrieveFrequency procedure. Then $H_i$ and $H'_i$ are complements in $\mathcal{G}_i$.

**Lemma B.1.** *Let $L' = L + 1$. Fix $\epsilon, \rho \in (0, 1/2)$. Let $Z^* \subset [\Delta]^d$ be a set of optimal $k$-centers for the $k$-median problem of the input point set. For each $i \in [0, L]$, if at most $ekL'/\rho$ cells $\mathcal{C}$ in $\mathcal{G}_i$ satisfy $d(\mathcal{C}, Z^*) \leq \Delta/(2^{i+1}d)$, then with probability $1 - \rho/L'$, the following two statements hold:*

1. $\left|\sum_{\mathcal{C} \in H_i}(\widehat{|\mathcal{C}|} - |\mathcal{C}|)\left(d(c(\mathcal{C}), Z) - d(c(\mathcal{C}^P), Z)\right)\right| \leq \frac{\epsilon \mathsf{OPT}}{2L'}$ *for every* $Z \subset [\Delta]^d$.

2. $\sum_{\mathcal{C} \in H'_i} |\mathcal{C}| \mathrm{diam}(\mathcal{C}) \leq \beta \mathsf{OPT}$ *for $\beta = 3d^{3/2}$*

*Proof of Lemma B.1.* Let $L' = L + 1$. Fix a value $i \in [0, L]$ and then $W = \Delta/2^i$ is the width of a cell in $\mathcal{G}_i$. Since at most $ekL'/\rho$ cells in $\mathcal{G}_i$ satisfy $d(\mathcal{C}, Z^*) \leq W/(2d)$, then of the remaining cells, at most $2kL'/\rho$ cells can contain more than $\rho d\mathsf{OPT}/(WkL')$ points. This is because each such cells contribute at least $\frac{\rho d \mathsf{OPT}}{WkL'} \frac{W}{2d} = \frac{\rho \mathsf{OPT}}{2kL'}$ to the cost which sums to $\mathsf{OPT}$. Therefore, at most $(e+2)L'k/\rho$ cells contain more than $\rho d\mathsf{OPT}/(WkL')$ points.

The number of cells in grid $\mathcal{G}_i$ is at most $N = (1 + 2^i)^d$ (and perhaps as few as $2^{id}$, depending on the random vector $v$), so HEAVY-HITTER receives cells of at most $N$ types. Enumerating all cells $C \in \mathcal{G}_i$ such that $|C_j| \geq |C_{j+1}|$, define $f_j = |C_j|$. Algorithm 1 sets $k' = (e+2)L'k/\rho$, and the additive error of the estimator of $f_i$ of HEAVY-HITTER is given by $\epsilon' \sqrt{\sum_{j=k'+1}^N f_j^2}$. We know that for all $j > k'$ the value $f_j \leq \rho d\mathsf{OPT}/(WkL')$. Moreover, the sum $\sum_{j=k'+1}^N f_j \leq 2d\mathsf{OPT}/W$ because each point is at distance at least $W/(2d)$ to a point of $Z^*$. Under these two restraints, the grouping of maximal error is with $f_j = \rho d\mathsf{OPT}/(WkL')$ for $k' < j \leq k' + 2kL'/\rho$ and $f_j = 0$ for $j > k' + 2kL'/\rho$. Then the additive error becomes $\epsilon' \sqrt{2\rho/(kL')} d\mathsf{OPT}/W$.

The error from a single cell $\mathcal{C}_j$ is at most $|f_j - \hat{f}_j| \sqrt{d} W$, and HEAVY-HITTER gaurantees with probability $1 - \delta$ that $|f_j - \hat{f}_j| \leq \epsilon' \sqrt{2\rho/(kL')} d\mathsf{OPT}/W$ for every $j$. Therefore to ensure total error over all $k'$ cells is bounded by $\epsilon \mathsf{OPT}/(2L')$, we set $\epsilon' \leq \epsilon \sqrt{\frac{\rho}{8(2+e)^2 kd^3 L'^3}}$. Setting $\delta = \rho/L'$, the above bound holds with probability at least $1 - \rho/L'$.

For the second claim, we must bound $\sum_{\mathcal{C} \in H'_i} |\mathcal{C}|$. $H_i$ consists of the top $k'$ cells when ordered by value of $\hat{f}_j$. This may differ from the top $k'$ cells when ordered by value of $f_j$, but if $j$ and $j'$ change orders between these two orderings then $|f_j - f_{j'}| \leq 2\epsilon' \sqrt{2\rho/(kL')} d\mathsf{OPT}/W$. Since the sum may swap up to $k'$ indices, the difference is bounded by $2k'\epsilon' \sqrt{2\rho/(kL')} d\mathsf{OPT}/W$. By setting $\epsilon' \leq \sqrt{\frac{\rho}{8(2+e)^2 dkL'}}$, we can ensure that the difference is at most $d\mathsf{OPT}/W$. We know that $\sum_{j=k'+1}^N f_j \leq 2d\mathsf{OPT}/W$, and so $\sum_{\mathcal{C} \in H'_i} |\mathcal{C}| \leq 3d\mathsf{OPT}/W$. For all cells $\mathcal{C} \in \mathcal{G}_i$, $\mathrm{diam}(\mathcal{C}) = \sqrt{d} W$. Therefore $\sum_{\mathcal{C} \in H'_i} |\mathcal{C}| \mathrm{diam}(\mathcal{C}) \leq 3d^{3/2} \mathsf{OPT}$.

$\square$

**Lemma B.2.** *Let $L' = L + 1$. In Algorithm 1, fixing $\epsilon, \rho \in (0, 1/2)$, $o \in O$ and $i \in [0, L]$, if $\mathsf{OPT}/2 \leq o \leq \mathsf{OPT}$, then with probability $1 - \rho/(L'\Delta^{kd})$, at most $\frac{(2+e)L'k}{\rho} + \frac{24d^4 L'^3 k}{\epsilon^2} \ln \frac{1}{\rho}$ cells of $\mathcal{G}_i$ contain a point of $S_{i,o}$.*

*Proof.* Similar to the proof of Lemma B.1, there are at most $k' = (2+e)L'k/\rho$ cells $\mathcal{C}$ in $\mathcal{G}_i$ that satisfy $|\mathcal{C}| \geq \rho d\mathsf{OPT}/(Wk)$ and/or $d(\mathcal{C}, Z^*) \leq W/(2d)$. Considering the other cells, together they contain at most $2d\mathsf{OPT}/W$ points. So by a Chernoff bound, with probability $1 - \rho/(L'\Delta^{kd})$ at most $O(2d\pi_{i,o}\mathsf{OPT}/(W\rho)) \leq 24d^4L'^3 k \ln\frac{1}{\rho}/\epsilon^2$ points are sampled. The claim follows since each non-empty cell must contain at least one point. $\square$

*Proof of Theorem 3.6.* Let $L' = L + 1$. W.l.o.g. we assume $\rho \geq \Delta^{-d}$, since otherwise we store the entire set of points and the theorem is proved. By Lemma 2.2, with probability at least $1 - \rho$, for every level $i \in [0, L]$, at most $ekL'/\rho$ cells $\mathcal{C}$ in $\mathcal{G}_i$ satisfy $d(\mathcal{C}, Z^*) \leq \Delta/(2^{i+1}d)$. Conditioning on this event, we will show 1) in the query phase, if $o^* \leq \mathsf{OPT}$, then with probability at least $1 - 4\rho$, $S$ is the desired coreset; 2) there exists $o \leq \mathsf{OPT}$ in the guesses $O = \{1, 2, 4, \ldots, \Delta^{d+1}\}$ such that with probability $1 - 4\rho$, none of the $K$-set structures return Nil. 1) and 2) guarantee the correctness of the algorithm. Note that one can always rescale $\rho$ to $\rho/9$ to achieve the correct probability bound. Finally, we will bound the space, update time and query time of the algorithm.

To show 1), we first note that the coreset size is at most $O(KL)$ as desired. Then by Lemma 3.2, we only need to show that with probability at least $1 - 4\rho$, for any $k$-set $Z \subset [\Delta]^d$ and any level $i \in [-1, L]$,

$$|\mathrm{cost}(\mathcal{G}_i, Z) - \widehat{\mathrm{cost}}(\mathcal{G}_i, Z)| \leq \frac{\epsilon \mathsf{OPT}}{L'},$$

where the value of each $\widehat{|\mathcal{C}|}$ is returned by RetrieveFrequency. For each level $i$, we denote $C_i$ as the set of cells that gets frequency from a HEAVY-HITTER instances in the RetrieveFrequency procedure, and $S_i = \{p \in \mathcal{C} : \mathcal{C} \notin C_i, h_{o^*,i}(p) = 1\}$ be the set of points sampled in the rest of cells. Since $KS_{o^*,i}$ does not return Fail, then for each $\mathcal{C} \in \mathcal{G}_i \setminus C_i$, $\widehat{|\mathcal{C}|} = |S_i \cap \mathcal{C}|/\pi_i(o^*)$. Fix a $k$-set $Z \subset [\Delta]^d$, we rewrite the cost as,

$$\widehat{\mathrm{cost}}(\mathcal{G}_i, Z) = \sum_{\mathcal{C} \in C_i} \widehat{|\mathcal{C}|}\left(d(c(\mathcal{C}), Z) - d(c(\mathcal{C}^P), Z)\right) + \sum_{p \in S_i}(d(c_p^i, Z) - d(c_p^{i-1}, Z)))/\pi_i(o^*),$$

where $\mathcal{C}^P$ is the parent cell of $\mathcal{C}$ in grid $\mathcal{G}_{i-1}$. By Lemma B.1 we have that, with probability at least $1 - \rho/L'$, for every $Z \subset [\Delta]^d$,

$$\left|\sum_{\mathcal{C} \in C_i} \widehat{|\mathcal{C}|}\left(d(c(\mathcal{C}), Z) - d(c(\mathcal{C}^P), Z)\right) - \sum_{\mathcal{C} \in C_i} |\mathcal{C}|\left(d(c(\mathcal{C}), Z) - d(c(\mathcal{C}^P), Z)\right)\right| \leq \frac{\epsilon \mathsf{OPT}}{2L'},$$

and that, $\sum_{\mathcal{C} \in \mathcal{G}_i \setminus C_i} |\mathcal{C}|\mathrm{diam}(\mathcal{C}) \leq 3d^{3/2}\mathsf{OPT}$. Conditioning on this event, by Lemma 3.4, with probability at least $1 - \rho/(L'\Delta^{kd})$,

$$\left|\sum_{p \in S_i}(d(c_p^i, Z) - d(c_p^{i-1}, Z)))/\pi_i - \sum_{\mathcal{C} \in \mathcal{G}_i \setminus C_i} |\mathcal{C}|\left(d(c(\mathcal{C}), Z) - d(c(\mathcal{C}^P), Z)\right)\right| \leq \frac{\epsilon \mathsf{OPT}}{2L'}.$$

By a union bound, we show with probability at least $1 - 4\rho$, for any $k$-set $Z \subset [\Delta]^d$ and any level $i \in [-1, L]$,

$$|\mathrm{cost}(\mathcal{G}_i, Z) - \widehat{\mathrm{cost}}(\mathcal{G}_i, Z)| \leq \frac{\epsilon \mathsf{OPT}}{L'},$$

as desired.

To show 2), we will consider some $\text{OPT}/2 \leq o \leq \text{OPT}$. By Lemma B.2 with probablity at least $1 - \rho/\Delta^{kd}$, the total number of cells occupied by sample points in each level is upper bounded by $K = \frac{(2+e)L'k}{\rho} + \frac{24d^4L'^3k}{\epsilon^2} \ln \frac{1}{\rho}$. Thus by the guarantee of the K-Set structure, with probability at least $1 - \rho$, none of the $\text{KS}_{o,0}, \text{KS}_{o,1} \ldots, \text{KS}_{o,L}$ will return Fail.

The memory requirement of the algorithm is determined by the $L$ instances of HEAVY-HITTER and the $dL^2$ instances of $K$-set. By Theorem 3.5, each instance of HEAVY-HITTER requires $O\left((k' + \frac{1}{\epsilon'^2}) \log \frac{N}{\delta} \log m\right)$ bits of space. Here $N \leq (1 + \Delta/W)^d \leq \Delta^d$ and $m$ is the maximum number of elements active in the stream. Since we require that at most one point exists at each location at the same time, we have that $m \leq N$. The parameters are set to $k' = (2+e)Lk/\rho$, $\epsilon' = \left(\epsilon \sqrt{\frac{\rho}{8(2+e)^2 kd^3L^3}}\right)$, and $\delta = \rho/L$. This translates to a space bound of $O\left(dL + \log \frac{1}{\rho}\right) \frac{d^4 L^5 k}{\rho \epsilon^2}$ bits. For each K-Set data structure, it requires

$$O(KdL \log(KL/\rho)) = O\left(\frac{d^5 L^4 k}{\epsilon^2} + \frac{dkL^2}{\rho}\right) \log \frac{dkL}{\epsilon \rho}$$

bits of space. In total, there are $O(dL^2)$ K-Set instances and thus all K-Set instances cost $O\left(\frac{d^6 L^6 k}{\epsilon^2} + \frac{d^2 kL^4}{\rho}\right) \log \frac{dkL}{\epsilon \rho}$ bits of space. By the same argument as in the offline case, the last paragraph of the proof of Theorem 3.3, the size of the coreset is at most $O((k' + K)L) = O(d^4 kL^4 \epsilon^{-2} + kL^2/\rho)$ points. Finally, to derandomize the fully random functions, we use Nissan's pseudo-random generator [Nis92] in a similar way used in [Ind00b]. But our pseudo-random bits only need to fool the sampling part of the algorithm rather than whole algorithm. We consider an augmented streaming algorithm $\mathcal{A}$ that does exactly the same as in CoreSet but with all the HEAVY-HITTER operations removed. Thus all $K$-set instances will have identical distribution with the ones in algorithm CoreSet. $\mathcal{A}$ uses $O(KdL \log(KL/\rho))$ bits of space. To fool this algorithm, using Nissan's pseudo-random generator, the length of random seed to generate the hash functions we need is of size $O(KdL \log(KL/\rho) \log(|O|\Delta^d)) = O\left(\left(\frac{d^7 kL^7}{\epsilon^2} + \frac{d^3 kL^5}{\rho}\right) \log \frac{dkL}{\rho \epsilon}\right)$. This random seed is thus sufficient to be used in Algorithm CoreSet. Thus the total space used in the algorithm is $O\left(\left(\frac{d^7 kL^7}{\epsilon^2} + \frac{d^3 kL^5}{\rho}\right) \log \frac{dkL}{\rho \epsilon} + \frac{d^5 kL^6}{\epsilon^2 \rho}\right)$ bits.

Regarding the update time, for the HEAVY-HITTER operations, it requires $O(L \log N) = O(dL^2)$ time. For the $K$-set operations, it requires $|O|LO(\log(KL/\rho)) = dL^2 \log(dkL/(\rho\epsilon))$ time. The derandomized hash operation takes $O(dL)$ more time per update. The final query time is dominated by the HEAVY-HITTER data structure, which requires $\text{poly}(d, k, L, 1/\epsilon)$ time. $\square$

## C  Full Construction of Positively Weighted Coreset

In this section, we will introduce a modification to our previous coreset construction, which leads to a coreset with all positively weighted points. The high level idea is as follows. When considering the estimate of the number of points in a cell, the estimate is only accurate when it truly contains a large number of points. However, in the construction of the previous section, we sample from each cell of each level, even though some of the cells contain a single point. For those cells, we cannot adjust their weights from negative to positive, since doing so would introduce large error. In this section, we introduce an ending level to each point. In other words, the number of points of a cell is estimated by sampling only if it contains many points. Thus, the estimates will be accurate enough and allow us to rectify the weights to be all positive.

This section is organized as follows. We reformulate the telescope sum in Subsection 4.1, provide a different construction (still with negative weights) in Subsection 4.2, modify our different construction to output non-negative weights in Subsection 4.3, and move this construction into to the streaming setting in Subsection 4.4. For simplicity of presentation, we will use $\lambda_1, \lambda_2, \ldots$ to denote some fixed positive universal constants.

## C.1 Reformulation of the Telescope Sum

**Definition C.1.** *A* heavy cell identification scheme $\mathcal{H}$ *is a map* $\mathcal{H} : \mathcal{G} \to \{heavy, non\text{-}heavy\}$ *such that,* $h(\mathcal{C}_{-1}) =$ *heavy and for cell* $\mathcal{C} \in \mathcal{G}_i$ *for* $i \in [0, L]$

1. *if* $|\mathcal{C}| \geq \frac{2^i \rho d \mathsf{OPT}}{k(L+1)\Delta}$ *then* $\mathcal{H}(\mathcal{C}) =$ *heavy;*

2. *If* $\mathcal{H}(\mathcal{C}) =$ *non-heavy, then* $\mathcal{H}(\mathcal{C}') =$ *non-heavy for every subcell* $\mathcal{C}'$ *of* $\mathcal{C}$.

3. *For every cell* $\mathcal{C}$ *in level* $L$, $\mathcal{H}(\mathcal{C}) =$ *non-heavy.*

4. *For each* $i \in [0, L]$, $|\{\mathcal{C} \in \mathcal{G}_i : \mathcal{H}(\mathcal{C}) = heavy\}| \leq \frac{\lambda_1 k L}{\rho}$, *where* $\lambda_1 \leq 10$ *is a positive universal constant.*

*The output for a cell not specified by the above conditions can be arbitrary. We call a cell* heavy *if it is identified heavy by* $\mathcal{H}$. *Note that a heavy cell does not necessarily contain a large number of points, but the total number of these cells is always bounded.*

In the sequel, heavy cells are defined by an arbitrary fixed identification scheme unless otherwise specified.

**Definition C.2.** *Fix a heavy cell identification scheme* $\mathcal{H}$. *For level* $i \in [-1, L]$, *let* $\mathcal{C}(p, i) \in \mathcal{G}_i$ *be the cell in* $\mathcal{G}_i$ *containing* $p$. *The* ending level $l(p)$ *of a point* $p \in P$ *is the largest level* $i$ *such that* $\mathcal{H}(\mathcal{C}(p, i)) =$ *heavy, and* $\mathcal{H}(\mathcal{C}(p, i+1)) =$ *non-heavy.*

Note that the ending level is uniquely defined if a heavy cell identification scheme is fixed. We now rewrite the telescope sum for $p$ as follows,

$$p = \sum_{i=0}^{l(p)} \left(c_p^i - c_p^{i-1}\right) + c_p^L - c_p^{l(p)},$$

where $c_p^{-1} = \mathbf{0}$ and $c_p^L = p$. For arbitrary $k$-centers $Z \subset [\Delta]^d$, we write,

$d(p, Z) = \sum_{i=0}^{l(p)} \left(d(c_p^i, Z) - d(c_p^{i-1}, Z)\right) + d(c_p^L, Z) - d(c_p^{l(p)}, Z) + d(\mathbf{0}, Z)$

Let $P_l$ be all the points with ending level $l(p) = l$. We now present the following lemmas.

**Lemma C.3.** *Let* $P_i$ *be the set of points with ending level* $i$. *Let* $Z^* \subset [\Delta]^d$ *be a set of optimal $k$-centers for the $k$-median problem of the input point set. Assume that for each* $i \in [-1, L]$, *at most* $ek(L+1)/\rho$ *cells* $\mathcal{C}$ *in* $\mathcal{G}_i$ *satisfy* $d(\mathcal{C}, Z^*) \leq \Delta/(2^{i+1}d)$. *Then*

$$|P_i| \cdot \frac{\Delta \sqrt{d}}{2^i} \leq \lambda_2 d^{3/2} \mathsf{OPT},$$

*where* $\lambda_2 > 0$ *is a universal constant.*

Before we prove this lemma, we first introduce the following lemmas to bound the cells with a large number of points.

**Lemma C.4.** *Assume that for each $i \in [0, L]$, at most $ek(L+1)/\rho$ cells $\mathcal{C}$ in $\mathcal{G}_i$ satisfy $d(\mathcal{C}, Z^*) \leq \Delta/(2^{i+1}d)$. Then for any $r > 0$ there are at most $\frac{(e+2r)k(L+1)}{\rho}$ cells that satisfy $|\mathcal{C}| \geq \frac{2^i \rho d \mathsf{OPT}}{rk(L+1)\Delta}$.*

*Proof of Lemma C.4.* Let $L' = L + 1$. Fix a value $i \in [0, L]$ and then $W = \Delta/2^i$ is the width of a cell in $\mathcal{G}_i$. Since at most $ekL/\rho$ cells in $\mathcal{G}_i$ satisfy $d(\mathcal{C}, Z^*) \leq W/(2d)$, then of the remaining cells, each contribute at least $\frac{\rho d \mathsf{OPT}}{rWkL'} \cdot \frac{W}{2d} = \frac{\rho \mathsf{OPT}}{2rkL'}$ to the cost, and the cost of these cells is at most $\mathsf{OPT}$. Therefore there can be at most $2rL'k/\rho$ cells such that $d(\mathcal{C}, Z^*) > W/(2d)$. Along with the at most $ekL'/\rho$ cells (by the assumption) such that $d(\mathcal{C}, Z^*) \leq W/(2d)$, there are at most $(e+2r)L'k/\rho$ cells that contain at least $\rho d \mathsf{OPT}/(rWkL')$ points. □

**Lemma C.5.** *Assume that for each $i \in [0, L]$, at most $ek(L+1)/\rho$ cells $\mathcal{C}$ in $\mathcal{G}_i$ satisfy $d(\mathcal{C}, Z^*) \leq \Delta/(2^{i+1}d)$. Then for $i \in [-1, L]$, the points of $P_i$ can be partitioned to at most $k' = \frac{2(e+6)k(L+1)}{\rho}$ groups, $G_1, G_2, \ldots, G_{k'}$, such that for each $j \in [k']$, there exists a $\mathcal{C} \in \mathcal{G}_i$, such that $G_j \in \mathcal{C}$, $|G_j| < 5 \frac{2^{i-1} \rho d \mathsf{OPT}}{k(L+1)\Delta}$.*

*Proof of Lemma C.5.* Let $L' = L + 1$. For each heavy cell in $\mathcal{G}_i$, if the number of points falling into its non-heavy subcells (in $\mathcal{G}_{i+1}$) is less than $\frac{2^{i-1} \rho d \mathsf{OPT}}{kL'\Delta}$, we group all these subcells into a single group. Let the groups formed this way be called type I, and by Property 4 of Definition 4.1 there are at most $(e+4)kL'/\rho$ type I groups.

For each of the remaining heavy cells in $\mathcal{G}_i$, we group its subcells into groups such that each group contains a number of points in the interval $\left[\frac{2^{i-1}\rho d \mathsf{OPT}}{kL'\Delta}, 5\frac{2^{i-1}\rho d \mathsf{OPT}}{kL'\Delta}\right)$. This can be done since each non-heavy subcell contains less than $\frac{2^{i+1}\rho d \mathsf{OPT}}{kL'\Delta} = 4\frac{2^{i-1}\rho d \mathsf{OPT}}{kL'\Delta}$ points, and the total number of points contained in them is at least $\frac{2^{i-1}\rho d \mathsf{OPT}}{kL'\Delta}$ (otherwise we would have formed a type I group). Let the groups formed this way be called type II. By the assumption of of Lemma C.3, at most $\frac{ekL'}{\rho}$ of these non-heavy subcells are within distance $\frac{\Delta}{2^{i+2}d}$ from an optimal center of $Z^*$. Since each type II group contains at least $\frac{2^{i-1}\rho d \mathsf{OPT}}{kL'\Delta}$ points, by the same argument as in Lemma C.4, the number of type II groups further than distance $\frac{\Delta}{2^{i+2}d}$ from an optimal center is at most $\frac{8kL'}{\rho}$. We conclude that,

$$k' \leq \frac{(e+4)kL'}{\rho} + \frac{(e+8)kL'}{\rho}.$$

□

*Proof of Lemma C.3.* Let $L' = L + 1$. Fix a value $i \in [-1, L]$ and then $W = \Delta/2^i$ is the upper bound of the width of a cell in $\mathcal{G}_i$. Let $G_1, G_2, \ldots G_{k'}$ be group of points satisfying Lemma C.5. Thus, $\sum_{p \in P_l} \frac{\Delta\sqrt{d}}{2^i} \leq \sum_{j \in [k']} \frac{2^{i+1}\rho d \mathsf{OPT}}{kL'\Delta} \cdot \frac{\Delta\sqrt{d}}{2^i} \leq \frac{\lambda' kL'}{\rho} \cdot \frac{2^{i+1}\rho d \mathsf{OPT}}{kL'\Delta} \frac{\Delta\sqrt{d}}{2^i} \leq \lambda_2 d^{3/2} \mathsf{OPT}$ for some universal constants $\lambda'$ and $\lambda_2$. □

*Proof of Proposition C.10.* First notice that the weighted set satisfies the about condition is an $\epsilon$-coreset. If we replace each $|\widehat{\mathcal{C}_i}|$ by the exact number of points in $|\mathcal{C}_i|$, then the new weighted set is an $(\epsilon/2)$-coreset. For each $\mathcal{C} \in \mathcal{G}$, let $b_\mathcal{C}$ be the new value returned by the algorithm, and $b_q$ is the new value of a point $q \in S$. The error of the cost introduced is at most,

$$A = \sum_{i=0}^{L} \left( \sum_{\mathcal{C} \in \mathcal{G}_i: \text{ heavy}} ||\widehat{\mathcal{C}}| - b_\mathcal{C}| + \sum_{p \in S_{i-1}} \left|\left(\frac{1}{\pi_{i-1}} - b_p\right)\right| \right) \frac{\Delta\sqrt{d}}{2^i}.$$

By the procedure, the new value of a cell is always smaller than its original value, thus

$$A = \sum_{i=0}^{L} \left( \sum_{\mathcal{C} \in \mathcal{G}_i: \text{ heavy}} |\widehat{\mathcal{C}}| - b_{\mathcal{C}} + \sum_{p \in S_{i-1}} \left( \frac{1}{\pi_{i-1}} - b_p \right) \right) \frac{\Delta \sqrt{d}}{2^i} = \sum_{i=0}^{L} g_i,$$

where

$$g_i = \left( \sum_{\mathcal{C} \in \mathcal{G}_i: \text{heavy}} |\widehat{\mathcal{C}}| - b_{\mathcal{C}} + \sum_{p \in S_{i-1}} \frac{1}{\pi_{i-1}} - b_p \right) \frac{\Delta \sqrt{d}}{2^i}.$$

Let

$$f_i = \left( \sum_{\mathcal{C} \in \mathcal{G}_i: \text{heavy}} \left| |\mathcal{C}| - |\widehat{\mathcal{C}}| \right| + \sum_{\mathcal{C}' \in \mathcal{G}_{i-1}: \text{heavy}} \left| \frac{|S_{i-1} \cap \mathcal{C}'|}{\pi_{i-1}} - |P_{i-1} \cap \mathcal{C}'| \right| \right) \frac{\Delta \sqrt{d}}{2^i}.$$

Thus $f_i \leq \epsilon \mathsf{OPT}/L$ by choosing appropriate $\lambda_6$. Now consider heavy cell $\mathcal{C} \in \mathcal{G}_i$, let $s_{\mathcal{C}} = \left| b_{\mathcal{C}} - \sum_{\mathcal{C} \in \mathcal{G}_{i+1}: \text{heavy}} |\widehat{\mathcal{C}}| - \frac{|S_i \cap \mathcal{C}|}{\pi_i} \right|$. Then,

$$s_{\mathcal{C}} = \left| b_{\mathcal{C}} - |\widehat{\mathcal{C}}| + |\widehat{\mathcal{C}}| - |\mathcal{C}| - \sum_{\mathcal{C}' \in \mathcal{G}_{i+1}: \text{heavy}} (|\widehat{\mathcal{C}'}| - |\mathcal{C}'|) - \left( \frac{|S_i \cap \mathcal{C}|}{\pi_i} - |P_i \cap \mathcal{C}| \right) \right|$$

$$\leq \left| b_{\mathcal{C}} - |\widehat{\mathcal{C}}| \right| + \left| |\widehat{\mathcal{C}}| - |\mathcal{C}| \right| + \sum_{\mathcal{C}' \in \mathcal{G}_{i+1}: \text{heavy}} \left| |\widehat{\mathcal{C}'}| - |\mathcal{C}'| \right| + \left| \frac{|S_i \cap \mathcal{C}|}{\pi_i} - |P_i \cap \mathcal{C}| \right|. \quad (9)$$

Then

$$g_i = \sum_{\mathcal{C} \in \mathcal{G}_{i-1}} s_{\mathcal{C}} \frac{\Delta \sqrt{d}}{2^i} + \left( \sum_{p \in S_{i-1}} \frac{1}{\pi_{i-1}} - b_p \right) \frac{\Delta \sqrt{d}}{2^i} \leq \frac{1}{2} g_{i-1} + \frac{1}{2} f_{i-1} + f_i. \quad (10)$$

Since $g_{-1} = f_{-1} = 0$, thus

$$g_i \leq f_i + 3 \sum_{j=0}^{i-1} 2^{j-i} f_j, \text{ and } \sum_{i=0}^{L} g_i \leq \sum_{i=0}^{L} f_i (1 + \sum_{j=1}^{i} \frac{3}{2^j}) \leq 4 \sum_{i=1}^{L} f_i \leq 4\epsilon \mathsf{OPT}. \qquad \square$$

**Remark C.6.** *The multiset of centers of heavy cells with each assigned a weight of the number of points in the cell is a $O(d^{3/2})$-coreset. This can be easily seen by removing the term of $d(c_p^L, Z) - d(c_p^{l(p)}, Z)$ from Equation (C.1) together with Lemma C.3, which bounds the error introduced by this operation.*

### C.2 The New Construction (with arbitrary weights)

For these heavy cells, we use `HEAVY-HITTER` algorithms to obtain accurate estimates of the number of points in these cells, thus providing a *heavy cell identification scheme*. For the non-heavy cells, we only need to sample points from the bottom level, $\mathcal{G}_L$, but with a different probability for points with different ending levels. We present the following lemma that governs the correctness of sampling from the last level.

**Lemma C.7.** *If a set of points $P_i \subset P$ satisfies $|P_i|\Delta\sqrt{d}/(2^i) \leq \beta\text{OPT}$ for some $\beta \geq 2\epsilon/(3(L+1))$, let $S_i$ be an independent sample from $P_i$ such that $p \in S_i$ with probability*

$$\pi_i \geq \min\left(\frac{3a(L+1)^2\Delta\sqrt{d}\beta}{2^i\epsilon^2 o}\ln\frac{2\Delta^{kd}(L+1)}{\rho}, 1\right)$$

*where $0 < o \leq a\text{OPT}$ for some $a > 0$. Then for a fixed set $Z \subset [\Delta]^d$, with probability at least $1 - \rho/((L+1)Delta^{kd})$, $|\sum_{p\in S_i}(d(c_p^i, Z) - d(p,Z))/\pi_i - \sum_{p\in P_i}(d(c_p^i, Z) - d(p,Z))| \leq \frac{\epsilon\text{OPT}}{L+1}$.*

*Proof.* The proof is identical to that of Lemma 3.4. □

**Lemma C.8.** *Consider a set of sets $\{P_i\}_{i=0}^L$ which satisfies $|P_i|\Delta\sqrt{d}/(2^i) \leq \frac{\beta\rho}{k(L+1)}\text{OPT}$ for some $\beta \geq \epsilon/(3(L+1))$. For each $i \in [0, L]$, let $S_i$ be an independent sample from $P_i$ with sampling probability*

$$\pi_i \geq \min\left(\frac{4a\beta k(L+1)^3\Delta\sqrt{d}}{2^i\epsilon^2\rho o}\log\frac{2}{\delta}, 1\right)$$

*where $0 < o \leq a\text{OPT}$ for some $a > 0$, then with probability at least $1 - \delta$,*

$$\left|\frac{|S_i \cap P_i|}{\pi_i} - |P_i|\right|\frac{\Delta\sqrt{d}}{2^i} \leq \frac{\epsilon\rho\text{OPT}}{k(L+1)^2}.$$

*Proof of Lemma C.8.* The proof is simply by Bernstein inequality. Let $t = \frac{2^i\epsilon\rho\text{OPT}}{\sqrt{d}k(L+1)^2\Delta}$, $X_p := \mathbb{I}_{p\in S_i}/\pi_i$, then we have that $Var(X_p) \leq 1/\pi_i$ and $b := \max_p |X_p| \leq 1/\pi_i$. By Bernstein's inequality, for any $j \in [k']$,

$$Pr\left[\left|\frac{|P_j \cap S_i|}{\pi_i} - |P_j|\right| > t\right] \leq 2e^{-\frac{t^2}{2|\mathcal{C}|/\pi_i + 2bt/3}} \leq \delta. \quad \square$$

We now describe the new construction. This essentially has the same gaurantee as the simpler construction from the previous section, however the benefit here is that (as shown in the next subsection) it can be modified to output only positive weights. In the following paragraph, the estimations $\widehat{|\mathcal{C}|}$ are given as a blackbox. In proposition C.9 we specify the conditions these estimations must satisfy.

**Non-Negatively Weighted Construction** Fix an arbitrary heavy cell identification scheme $\mathcal{H}$. Let $P_l$ be all the points with ending level $l(p) = l$. For each heavy cell $\mathcal{C}$, let $\widehat{|\mathcal{C}|}$ be an estimation of number of points of $|\mathcal{C}|$, we also call $\widehat{|\mathcal{C}|}$ the *value* of cell $\mathcal{C}$. For each non-heavy cell $\mathcal{C}'$, let $\widehat{|\mathcal{C}'|} = 0$. Let $S$ be a set samples of $P$ constructed as follows: $S = S_{-1} \cup S_0 \cup S_1 \cup \ldots \cup S_L$, where $S_l$ is a set of i.i.d samples from $P_l$ with probability $\pi_l$. Here $\pi_l$ for $l \in [-1, L]$ is redefined as, $\pi_l = \min\left(\frac{\lambda_3 d^2 \Delta L^2}{2^l\epsilon^2 o}\log\left(\frac{2L\Delta^{dk}}{\rho}\right) + \frac{\lambda_4 d^2 kL^3\Delta}{2^i\epsilon^2\rho o}\log\frac{30kL^2}{\rho^2}, 1\right)$ where $\lambda_3 > 0$ and $\lambda_4 > 0$ are universal constants. Our coreset $\mathcal{S}$ is composed by all the sampled points in $S$ and the cell centers of heavy cells, with each point $p$ assigned a weight $1/\pi_{l(p)}$ and for each cell center $c$ of a heavy cell $\mathcal{C} \in \mathcal{G}_i$, the weight is,

$$\text{wt}(c) = \widehat{|\mathcal{C}|} - \sum_{\substack{\mathcal{C}':\mathcal{C}'\in\mathcal{G}_{i+1},\mathcal{C}'\subset\mathcal{C},\\ \mathcal{C}'\text{ is heavy}}} \widehat{|\mathcal{C}'|} - \frac{|S_i \cap \mathcal{C}|}{\pi_i}. \tag{11}$$

For each non-heavy cell $\mathcal{C}$ except for those in the bottom level, $\text{wt}(c(\mathcal{C})) = 0$. The weight of each point from $S$ is the value of the corresponding cell in the bottom level.

We now state the following proposition for a coreset construction, which immediately serves as an offline coreset construction.

**Proposition C.9.** *Let $\mathcal{H}$ be an arbitrary heavy cell identification scheme. Fix $\Omega(\Delta^{-d}) \leq \rho < 1$ and for each heavy $\mathcal{C} \in \mathcal{G}_i$ in level $i$, $\widehat{|\mathcal{C}|}$ is an estimation of number of points in $\mathcal{C}$ with additive error at most $\frac{\epsilon}{\lambda_5 L d^{3/2}} \cdot \frac{2^i \rho d \text{OPT}}{kL\Delta}$, where $\lambda_5 > 0$ is a universal constant. Let $S_l$ be the set of i.i.d. samples of $P_l$ with probability $\pi_l(o)$. If $0 < o \leq \text{OPT}$, then with probability at least $1 - 4\rho$, for every $k$-set $Z \subset [\Delta]^d$,*

$$\left| \sum_{q \in S} \text{wt}(q) d(q, Z) - \sum_{p \in P} d(p, Z) \right| \leq \epsilon \text{OPT}.$$

*And the coreset size $|S|$ is*

$$O\left[ \frac{d^3 L^4 k}{\epsilon^2} \left( d + \frac{1}{\rho} \log \frac{kL}{\rho} \right) \frac{\text{OPT}}{o} \right].$$

*Proof of Proposition C.9.* Fix a $k$-set $Z \subset [\Delta]^d$. First notice that,

$$\widehat{\text{cost}}(Z) = \sum_{q \in S} \text{wt}(q) d(q, Z)$$

$$= \sum_{i=-1}^{L-1} \left[ \sum_{\mathcal{C} \in \mathcal{G}_i : \mathcal{C} \text{ heavy}} \left( \widehat{|\mathcal{C}|} - \sum_{\substack{\mathcal{C}' : \mathcal{C}' \in \mathcal{G}_{i+1}, \mathcal{C}' \subset \mathcal{C}, \\ \mathcal{C}' \text{ is heavy}}} \widehat{|\mathcal{C}'|} - \frac{|S_i \cap \mathcal{C}|}{\pi_i} \right) d(c(\mathcal{C}), Z) + \sum_{p \in S_i} \frac{d(p, Z)}{\pi_i} \right]$$

$$= \sum_{i=-1}^{L-1} \left[ \sum_{\mathcal{C} \in \mathcal{G}_i : \mathcal{C} \text{ heavy}} \widehat{|\mathcal{C}|} (d(c(\mathcal{C}), Z) - d(c(\mathcal{C}^P), Z)) + \sum_{p \in S_i} \frac{d(p, Z) - d(c_p^i, Z)}{\pi_i} \right], \quad (12)$$

where we denote $d(c(\mathcal{C}_{-1}^P), Z) = 0$ for convenience. Let $\text{cost}(Z) = \sum_{p \in P} d(p, Z)$. Note that we can also write the true cost of $Z$ as

$$\text{cost}(Z) = \sum_{i=-1}^{L-1} \left[ \sum_{\mathcal{C} \in \mathcal{G}_i : \mathcal{C} \text{ heavy}} |\mathcal{C}| (d(c(\mathcal{C}), Z) - d(c(\mathcal{C}^P, Z))) + \sum_{p \in P_i} d(p, Z) - d(c_p^i, Z) \right].$$

We have that,
$$\widehat{\text{cost}}(Z) - \text{cost}(Z) = A_1 + A_2,$$

where

$$A_1 = \sum_{i=-1}^{L-1} \left[ \sum_{\mathcal{C} \in \mathcal{G}_i : \mathcal{C} \text{ heavy}} (\widehat{|\mathcal{C}|} - |\mathcal{C}|)(d(c(\mathcal{C}), Z) - d(c(\mathcal{C}^P, Z))) \right]$$

and

$$A_2 = \sum_{i=-1}^{L-1} \left( \sum_{p \in S_i} \frac{d(p, Z) - d(c_p^i, Z)}{\pi_i} - \sum_{p \in P_i} d(p, Z) - d(c_p^i, Z) \right).$$

Let $Z^* \subset [\Delta]^d$ be a set of optimal $k$-centers for the $k$-median problem of the input point set. By Lemma 2.2, with probability at most $1 - \rho$, for each $i \in [0, L]$, if at most $ek(L+1)/\rho$ cells $\mathcal{C}$ in $\mathcal{G}_i$ satisfy $d(\mathcal{C}, Z^*) \leq \Delta/(2^{i+1}d)$. Conditioning on this event, we have that, by Lemma C.4 there are at most $k' = O\left(\frac{kL}{\rho}\right)$ heavy cells per level. Since for each $\mathcal{C} \in \mathcal{G}_i$, $\left|\widehat{|\mathcal{C}|} - |\mathcal{C}|\right| \leq \frac{\epsilon}{\lambda_5 L d^{3/2}} \cdot \frac{2^i \rho d \mathsf{OPT}}{kL\Delta}$, by choosing appropriate constant $\lambda_5 > 0$ we have

$$|A_1| \leq \sum_{i=-1}^{L-1} \left| \sum_{\mathcal{C} \in \mathcal{G}_i : \mathcal{C} \text{ heavy}} (\widehat{|\mathcal{C}|} - |\mathcal{C}|)(d(c(\mathcal{C}), Z) - d(c(\mathcal{C}^P), Z))) \right|$$

$$\leq L \cdot k' \cdot \frac{\epsilon}{\lambda_5 L d^{3/2}} \cdot \frac{2^i \rho d \mathsf{OPT}}{kL\Delta} \cdot \frac{\Delta\sqrt{d}}{2^i} \leq \frac{\epsilon \mathsf{OPT}}{2}. \tag{13}$$

For $A_2$, let

$$A_{2i} = \left( \sum_{p \in S_i} \frac{d(p, Z) - d(c_p^i, Z)}{\pi_i} - \sum_{p \in P_i} d(p, Z) - d(c_p^i, Z) \right).$$

By Lemma C.3, for each $i \in [-1, L-1]$, $|P_i|\Delta\sqrt{d}/(2^i) = \lambda_2(d^{3/2}\mathsf{OPT})$. Thus by Lemma C.7, and choosing appropriate constants, with probability at least $1 - \rho/(L+1)\Delta^{dk}$, $|A_{2i}| \leq \frac{\epsilon \mathsf{OPT}}{2(L+1)}$. By the union bound, with probability at least $1 - \rho$, for every level $i$, and every $k$-set $Z \subset [\Delta]^d$, $|A_{2i}| \leq \frac{\epsilon \mathsf{OPT}}{2(L+1)}$. Thus $|A_2| \leq \epsilon \mathsf{OPT}/2$. In total, with probability at least $1 - 3\rho$, $|A_1 + A_2| \leq \epsilon \mathsf{OPT}$ for any $k$-set $Z \subset [\Delta]^d$.

The coreset size is the number of heavy cells plus the number of sampled points. The number of heavy cells is $O(kL^2/\rho)$. The expected number of sampled points per level is at most,

$$|S_i| = O\left(\frac{d^4 L^3 k}{\epsilon^2} + \frac{d^3 L^2 k}{\epsilon^2 \rho} \log\left(\frac{kL}{\rho}\right)\right) \frac{\mathsf{OPT}}{o}.$$

By a Chernoff bound, with probability at least $1 - \rho/\Delta^{dk}$, for every level $i \in [0, L]$, the number of sampled points is,

$$|S_i| \leq O\left(\frac{d^4 L^3 k}{\epsilon^2} + \frac{d^3 L^3 k}{\rho \epsilon^2} \log\left(\frac{kL}{\rho}\right)\right) \frac{\mathsf{OPT}}{o}.$$

Thus the size of the coreset $\mathcal{S}$ is,

$$|\mathcal{S}| \leq O\left[\frac{d^3 L^4 k}{\epsilon^2} \left(d + \frac{1}{\rho} \log \frac{kL}{\rho}\right) \frac{\mathsf{OPT}}{o}\right].$$

$\square$

## C.3 Ensuring Non-Negative Weights

In this section, we will provide a procedure to rectify all the weights for the coreset constructed in the last sub-section. The idea is similar to the method used in [IP11]. The procedure is shown in Algorithm 4.3.

**Proposition C.10.** *Let $\mathcal{S}$ be a weighted set constructed using the Non-Negatively Weighted Construction, i.e. each heavy cell $\mathcal{C}$ has value $\widehat{|\mathcal{C}|}$ and the set of sampled points $S = S_{-1} \cup S_0 \ldots \cup S_L$* with each point in $S_l$ has weight $1/\pi_l$. If for each heavy cell $\mathcal{C} \in \mathcal{G}_i$, $||\widehat{\mathcal{C}}| - |\mathcal{C}|| \leq \frac{\epsilon}{\lambda_6 L d^{3/2}} \cdot \frac{2^i \rho d \mathsf{OPT}}{k L \Delta}$ for some universal constant $\lambda_6 > 0$ and for each $i \in [-1, L]$ and any $k$-set $Z \subset [\Delta]^d$,

$$\left| \sum_{p \in S} \mathrm{wt}(p)(d(c_p^i, Z) - d(p, Z)) - \sum_{p \in P_i} (d(c_p^i, Z) - d(p, Z)) \right|$$
$$\leq \frac{\epsilon \mathsf{OPT}}{2L},$$

and

$$\sum_{\mathcal{C} \in \mathcal{G}_i:\ heavy} \left| \frac{|S_i \cap \mathcal{C}|}{\pi_i} - |P_i \cap \mathcal{C}| \right| \frac{\Delta \sqrt{d}}{2^i} \leq \frac{\epsilon \mathsf{OPT}}{L}.$$

Then on input $|\widehat{\mathcal{C}_1}|, |\widehat{\mathcal{C}_2}|, \ldots, |\widehat{\mathcal{C}_{k'}}|$ and $S$, where $k'$ is the number of heavy cells, the coreset output by Algorithm 4.3 is a $(4\epsilon)$-coreset.

## C.4 The Streaming Algorithm

### C.4.1 Sampling From Sparse Cells

For the streaming algorithm, we can still use `HEAVY-HITTER` algorithms to find the heavy cells. The major challenge is to do the sampling for each point from its ending level. We do this using a combination of hash functions and `K-Set`. In Algorithm C.4.1, we provide a procedure that recovers the set of points from cells with a small number of points and ignore all the heavy cells. The guarantee is,

**Theorem C.11.** *Given as input a set of dynamically updating streaming points $P \subset [N]$, a set of mutually disjoint cells $C \subset [M]$, whose union covers the region of $P$. Algorithm C.4.1 outputs all the points in cells with less than $\beta$ points or output* `Fail`*. If with the promise that at most $\alpha$ cells from $C$ contain a point of $P$, then the algorithm outputs* `Fail` *with probability at most $\delta$. The algorithm uses $O(\alpha\beta(\log(M\beta) + \log N) \log N \log(\log N\alpha\beta/\delta) \log(\alpha\beta/\delta))$ bits in the worst case.*

The high level idea of this algorithm is to hash the original set of points to a universe of smaller size. For cells with less points, the collision rate is much smaller than cells with more points. To recover one bit of a point, we update that bit and the cell ID and also its hash tag to the `K-Set`-data structure. If there are no other points with hash values colliding with this point, the count of that point is simply 1. If this is the case, we immediately recover the bit. By repeating the above procedure once for each bit, we can successfully recover the set of points with no colliding hash tags. For those points with colliding hash tags, we simply ignore them. Each point has a constant probability to collide with another point, thus not be in the output. By running the whole procedure $O(\log(\alpha\beta/\epsilon))$ times in parallel, we reduce the probability to roughly $\epsilon$ for each point in cells with less than $\beta$ points. To formally prove Theorem C.11, we first prove the following lemma, which is the guarantee of Algorithm C.4.1.

**Lemma C.12.** *Given input a set of dynamically updating streaming points $P \subset [N]$, a set of mutually disjoint cells $C \subset [M]$, whose union covers the region of $P$. Algorithm C.4.1 outputs a set of points in cells with less than $\beta$ points or output* `Fail`*. If with the promise that at most $\alpha$ cells from $C$ contain a point of $P$, then the algorithm outputs* `Fail` *with probability at most* $\delta$. *Conditioning on the event that the algorithm does not output* `Fail`, *each point* $p$ *from cell with less than* $\beta$ *points is in the output with marginal probability at least* $0.9$. *The algorithm uses* $O(\alpha\beta(\log(M\beta) + \log N) \log N \log(\log N\alpha\beta/\delta))$ *bits in the worst case.*

*Proof.* We prove this lemma by showing that (a) if a point $p \in P$ contained in cell $\mathcal{C}$, with $|\mathcal{C} \cap P| \le \beta$, then with probability at least 0.99, there are no other points $p' \in \mathcal{C} \cap P$ with $H(p) = H(p')$, (b) conditioning on the event that the algorithm does not output `Fail`, then for any cell $\mathcal{C} \in C$, if a point $p \in \mathcal{C}$ such that no other points in $\mathcal{C} \cap P$ has the same hash value $H(p)$, then $p$ is in the output and (c) the algorithm outputs `Fail` with probability at most $\delta$. The correctness of the algorithm follows by (a), (b) and (c).

To show (a), consider any cell $\mathcal{C} \in C$ with $|\mathcal{C}| \le \beta$, let $p \in \mathcal{C}$ with hash value $H(p)$. Since $H$ is 2-wise independent, the expected number of other points hashed to the same hash value $H(p)$ is at most $\beta/U = 1/100$. By Markov's inequality, with probability at least 0.99, no other point in $\mathcal{C}$ is hashed to $H(p)$.

To show (b), notice that if the algorithm does not output `Fail`, then for a given cell $\mathcal{C}$, let $c$ be its ID, and $p_j$ be the $j$-th bit of point $p$. Then $(c, h, p_j)$ has 1 count and $(c, h, 1 - p_j)$ has 0 count for each $j \in [t]$, where $t = \lceil \log N \rceil$. Thus we can uniquely recover each bit of point $p$, hence the point $p$.

For (c), since there are at most $\alpha$ cells, there are at most $2\alpha U = O(\alpha\beta)$ many different updates for each `KS` structure. Therefore, with probability at most $\frac{\delta}{t}$, a single `KS` instance outputs `Fail`. By the union bound, with probability at least $1 - \delta$, no `KS` instance outputs `Fail`.

Finally, the space usage is dominated by the `KS` data structures. Since the input data to `KS` is from universe $[M] \times [U] \times \{0, 1\}$, each `KS` structure uses space $O(\alpha\beta(\log(M\beta) + \log N) \log(t\alpha\beta/\delta))$ bits of memory, the total space is $O(\alpha\beta(\log(M\beta) + \log N) \log N \log(t\alpha\beta/\delta))$. $\square$

*Proof of Theorem C.11.* Each instance of `SparseCellsSingle` fails with probability at most $\delta/(4A)$, where $A$ is the number independent `SparseCellsSingle` instances. By the union bound, with probability at least $1 - \delta/4$, none of them output `Fail`. Conditioning on this event, the random bits of the hash functions of each `SparseCellsSingle` instance are independent, thus by Lemma C.12 a fixed point $p \in \mathcal{C}$ with $|\mathcal{C}| \le \beta$ is in the output with probability at least $10^{-\log \frac{4\alpha\beta}{\delta}} \le \delta/(4\alpha\beta)$. Since there are at most $\alpha\beta$ points in cells with less than $\beta$ points, by the union bound we conclude that with probability at least $1 - \delta/4$, every point in cells with less than $\beta$ points is in the output set $S$. In sum, with probability at least $1 - \delta/2$, $S$ contains all the desired points.

The other `KS` instance outputs `Fail` with probability at most $\delta/2$. Thus if $T$ is not `Fail`, then $T$ contains the exact number of points of each cell. If any desired point is not in $S$, then $|\mathcal{C}| > |\mathcal{C} \cap S|$, we output `Fail`. This happens with probability at most $\delta$ under the gaurantee of the KSstructure.

Since each `SparseCellsSingle` instance uses $O(\alpha\beta(\log(M\beta) + \log N) \log N \log(tA\alpha\beta/\delta))$ bits of space, the final space of the algorithm is $O(\alpha\beta(\log(M\beta) + \log N) \log N \log(t\alpha\beta/\delta) \log(\alpha\beta/\delta))$. $\square$

### C.4.2 The Algorithm

With the construction of algorithm `SparseCells`, we now have all the tools for the streaming coreset construction. The streaming algorithm is composed by $O(L)$ levels of `HEAVY-HITTER` instances, which serve as a heavy cell identifier and by $O(L)$ levels of `SparseCells` instances, which sample the points from their ending levels. The full algorithm is stated in Algorithm 6. The guarantee of the algorithm is stated in the following theorem.

**Algorithm 4** SparseCells($N, M, \alpha, \beta, \delta$): input the point sets $P \subset [N]$ and set of cells $C \subset [M]$ such that at most $\alpha$ cells containing a point, output the set of points in cells with less than $\beta$ points.

Let $A \leftarrow \log \frac{4\alpha\beta}{\delta}$;
Let $R_1, R_2, \ldots, R_A$ be the results of independent instances of SparseCellsSingle($N, M, \alpha, \beta, \delta/(4A)$) running in parallel;
Let $T$ be the results of another parallel KS structure with space parameter $\alpha$ and error $\delta/2$ and with input as the cell IDs of points in $P$; /*$T$ returns the exact counts of each cell*/
**if** *any of the data structures returns* `Fail`:
| **return** `Fail`;
Let $S \leftarrow R_1 \cup R_2 \cup \ldots R_A$;
**if** $\exists$ *set* $\mathcal{C} \in T$ *with* $|\mathcal{C}| \leq \beta$ *and* $|\mathcal{C}| \neq |\mathcal{C} \cap S|$:
| **return** `Fail`;
**return** $S$;

---

**Theorem C.13.** *Fix $\epsilon, \rho \in (0, 1/2)$, positive integers $k$ and $\Delta$, Algorithm 6 makes a single pass over the streaming point set $P \subset [\Delta]^d$, outputs a weighted set $S$ with non-negative weights for each point, such that with probability at least $0.99$, $S$ is an $\epsilon$-coreset for $k$-median of size $O\left[\frac{d^3 L^4 k}{\epsilon^2}\left(d + \frac{1}{\rho} \log \frac{kL}{\rho}\right)\right]$, where $L = \log \Delta$. The algorithm uses*

$$O\left[\frac{d^7 L^7 k}{\epsilon^2}\left(\rho dL + \frac{1}{\rho}\log^2 \frac{dkL}{\rho\epsilon}(\log\log \frac{dkL}{\rho\epsilon} + L)\right)\log^2 \frac{dkL}{\rho\epsilon}\right]$$

*bits in the worst case. For each update of the input, the algorithm needs* $\text{poly}(d, 1/\epsilon, L, \log k)$ *time to process and outputs the coreset in time* $\text{poly}(d, k, L, 1/\epsilon, 1/\rho, \log k)$ *after one pass of the stream.*

*Proof.* W.l.o.g. assume $\rho \geq \Delta^{-d}$, since otherwise we can store the entire set of points. In the sequel, we will prove the theorem with parameter $O(\rho)$ and $O(\epsilon)$. It translates to $\rho$ and $\epsilon$ directly by scaling and with losing at most a constant factor in space and time bounds. By Lemma 2.2, with probability at least $1 - \rho$, for every level $i \in [0, L]$, at most $ekL/\rho$ cells $\mathcal{C}$ in $\mathcal{G}_i$ satisfy $d(C, Z^*) \leq \Delta/(2^{i+1}d)$. We condition on this event for the following proof.

We first show that the HEAVY-HITTER instances faithfully implement a *heavy cell identification scheme*. First note that with probability at least $1 - \rho$, all HEAVY-HITTER instances succeed. Conditioning on this event for the following proof. As shown in the proof of Lemma B.1, by setting $\epsilon' = \epsilon\sqrt{\frac{\rho}{\lambda_7 k d^3 L^3}}$ and $k' = \lambda_8 kL/\rho$, for appropriate positive universal constants $\lambda_7, \lambda_8$, then the additive error to each cell is at most $\frac{\epsilon}{\lambda_9 d^{3/2} L} \cdot \frac{2^i d\text{OPT}}{kL\Delta}$ for some universal constant $\lambda_9$, which matches the requirement of Proposition C.10. For each cell $\mathcal{C}$ with at least $2^i \rho d\text{OPT}/(k(L+1)\Delta)$ points, by Lemma C.4 it must be in the top $(e+2)k(L+1)/\rho$ cells. For each cell $\mathcal{C}'$ with at least $2^{i-1}\rho d\text{OPT}/(k(L+1)\Delta)$ points, it must be in the top $(e+4)k(L+1)/\rho$ cells. Since the additive error is $\frac{\epsilon}{\lambda_7 d^{3/2}(L+1)} \cdot \frac{2^i d\text{OPT}}{k(L+1)\Delta} \ll \frac{1}{2}\frac{2^i d\text{OPT}}{k(L+1)\Delta}$. Thus $\mathcal{C}$ is in the output of the HEAVY-HITTER instances, since otherwise $\widehat{|\mathcal{C}|} \leq \frac{1}{2}\frac{2^i d\text{OPT}}{k(L+1)\Delta} + \frac{\epsilon}{\lambda_7 d^{3/2}(L+1)} \cdot \frac{2^i d\text{OPT}}{k(L+1)\Delta}$ contradicts the error bound (by choosing sufficiently large $\lambda_7$). Thus the algorithm faithfully implements a heavy cell identification scheme.

Now we show that if there exists an $o \leq \text{OPT}$ such that no instance of SparseCells outputs Fail, then the result is a desired $O(\epsilon)$-coreset. This follows by Proposition C.9 and Proposition C.10. Then we note that the coreset size is upper bounded by $O\left[\frac{d^3 L^4 k}{\epsilon^2}\left(d + \frac{1}{\rho}\log \frac{kL}{\rho}\right)\right]$ as desired.

---

**Algorithm 5** SparseCellsSingle($N, M, \alpha, \beta, \delta$): input the point sets $P \subset [N]$ and set of cells $C \subset [M]$ such that at most $\alpha$ cells containing a point, output the set of points in cells with less than $\beta$ points.

---

**Initization**:
$U \leftarrow 100\beta$.
$t \leftarrow \lceil \log N \rceil$;
$H : [N] \rightarrow [U]$, 2-wise independent;
K-Set structures $\mathtt{KS}_1, \mathtt{KS}_2, \ldots \mathtt{KS}_t$ with space parameter $2\alpha U$ and probability $\frac{\delta}{t}$;

**Update**($p, op$):    /*$op \in \{\mathtt{Insert}, \mathtt{Delete}\}$*/
$c \leftarrow$ cell ID of $p$;
**for** $i \in [t]$:
    /*A point $p$ is represented as $(p_1, p_2, \ldots, p_t)$*/;
    $p_i \leftarrow$ the $i$-th bit of point $p$;
    $\mathtt{KS}_i.\text{update}((c, H(p), p_i), op)$;

**Query**:
**if** *for any* $i \in [t]$, $\mathtt{KS}_i$ *returns* Fail:
    **return** Fail;

$S \leftarrow \emptyset$;
**for** each $(c, h, p_1)$ in the output of $\mathtt{KS}_1$:
    **if** $(c, h, p_j) \notin \mathtt{KS}_j$ *for some* $j \in [t]$:
        /*A checking step, may not happen at all*/;
        **return** Fail;

    Let $s(c, h, p_j)$ be the counts of $(c, h, p_j)$ in $\mathtt{KS}_j$;
    **if** $s(c, h, p_j) = 1$ *and* $s(c, h, 1-p_j) = 0$ *for each* $j \in [t]$:
        $p \leftarrow (p_1, p_2, \ldots, p_t)$;
        $S \leftarrow S \cup \{p\}$;
**return** $S$

---

Next we show that there exists an $\mathtt{OPT}/2 \leq o^* \leq \mathtt{OPT}$ that with probability at least $1 - \rho$, no SparseCells instance $\mathtt{SC}_{o^*, i}$ outputs Fail. By Chernoff bound, with probability at least $1 - O(\rho)$, as also shown in the proof of Proposition C.9, per level at most $O\left[\frac{d^3 L^4 k}{\epsilon^2}\left(d + \frac{1}{\rho}\log\frac{kL}{\rho}\right)\frac{\mathtt{OPT}}{o}\right]$ cells is occupied by a point. And at most $O\left[\frac{d^3 L^2}{\epsilon^2}\left(\rho d + \log\frac{kL}{\rho} + \frac{\rho}{kL}\log\frac{L}{\rho}\right)\right]$ points is sampled per light cell. Conditioned on this fact and that each instances fails with probability at most $O(\rho/(dL))$, with probability at least $1 - O(\rho)$, no nstance $\mathtt{SC}_{o^*, i}$ fails.

Lastly, we bound the space usage and update/query time. For the HEAVY-HITTER instances, the total space used is $O\left(dL + \log\frac{1}{\rho}\right)\frac{d^4 L^5 k}{\rho\epsilon^2}$ bits, analogous to the proof of Theorem 3.6. Each SparseCells instance uses space $O\left[\frac{d^5 L^4 r^2 k}{\epsilon^2}\left(\rho dL + \frac{r^2(\log r + L)}{\rho}\right)\right]$, where $r = \log\frac{dkL}{\rho\epsilon}$. The total space bound is $O\left[\frac{d^6 L^6 r^2 k}{\epsilon^2}\left(\rho dL + \frac{r^2(\log r + L)}{\rho}\right)\right]$ bits. As a same argument in the proof of Theorem 3.6, the cost of de-randomization introduce an additional $dL$ factor. Thus, the final space bound is $O\left[\frac{d^7 L^7 r^2 k}{\epsilon^2}\left(\rho dL + \frac{r^2(\log r + L)}{\rho}\right)\right]$ bits. The query time and update time is similar to that of Theorem 3.6 thus poly $\left(d, L, \frac{1}{\epsilon}, \frac{1}{\rho}, k\right)$ and poly $\left(d, L, \frac{1}{\epsilon}, \log k\right)$.

□

## D  Synthetic Dataset

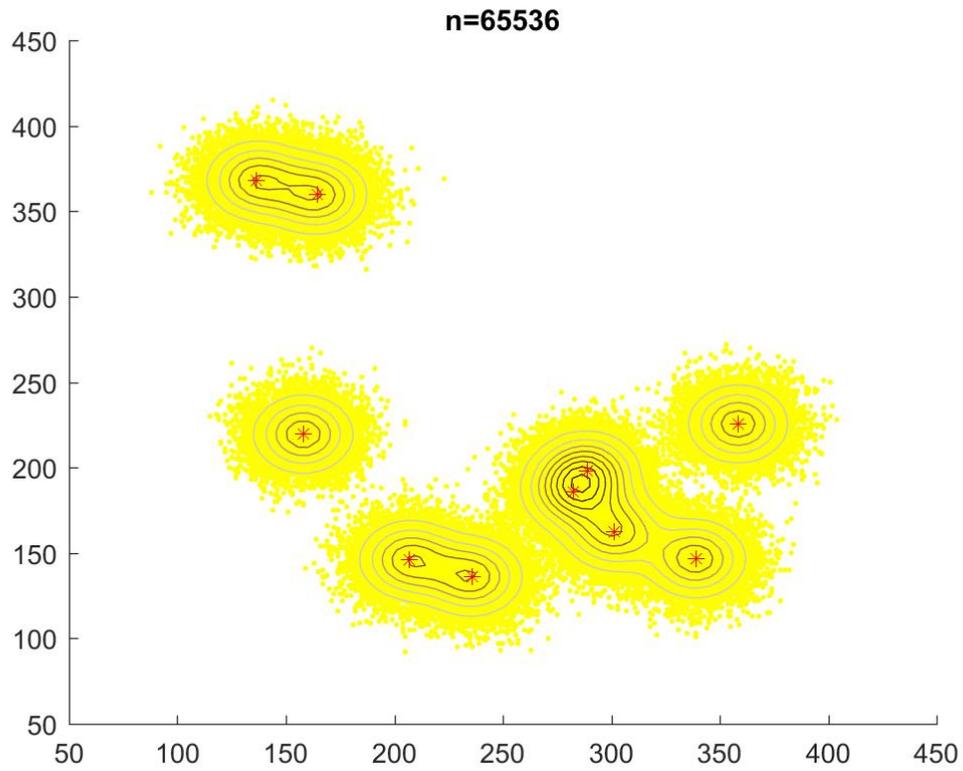

Figure 2: 65536 points are drawn from a Gaussian Mixture distribution. The contours illustrate the PDF function.

**Algorithm 6** PositiveCoreSet$(S, k, \rho, \epsilon)$: construct a $\epsilon$-coreset for dynamic stream $S$.

**Initization**:

Initialize a grid structure;

$O \leftarrow \{1, 2, 4, \ldots, \sqrt{d}\Delta^{d+1}\}$; $L \leftarrow \lceil \log \Delta \rceil$; $\pi_i(o) \leftarrow \min\left(\frac{\lambda_3 d^2 \Delta L^2}{2^l \epsilon^2 o} \log\left(\frac{2L\Delta^{dk}}{\rho}\right) + \frac{\lambda_4 d^2 kL^3 \Delta}{2^i \epsilon^2 \rho o} \log \frac{30kL^2}{\rho^2}, 1\right)$;

$\alpha \leftarrow O\left[\frac{d^3 L^4 k}{\epsilon^2}\left(d + \frac{1}{\rho} \log \frac{kL}{\rho}\right)\right]$, $\beta \leftarrow O\left[\frac{d^3 L^2}{\epsilon^2}\left(\rho d + \log \frac{kL}{\rho} + \frac{\rho}{kL} \log \frac{L}{\rho}\right)\right]$, $\epsilon' \leftarrow \epsilon\sqrt{\frac{\rho}{\lambda_7 kd^3 L^3}}$; $m \leftarrow 0$;

For each $o \in O$ and $i \in [0, L]$, construct fully independent hash function $h_{o,i} : [\Delta]^d \to \{0, 1\}$ with $Pr_{h_{o,i}}(h_{o,i}[q] = 1) = \pi_i(o)$; initialize SparseCells$(\Delta^d, (1+2^i)^d, \alpha, \beta, O(\rho/(dL)))$ instances $\text{SC}_{o,i}$;

Initialize HEAVY-HITTER$(\Delta^d, 10Lk/\rho, \epsilon', \rho/L)$ instances, $\text{HH}_0, \text{HH}_1, \ldots, \text{HH}_{L-1}$, one for a level;

**Update** $(S)$:

**for** *each update* $(op, q) \in S$:
    /*$op \in \{\text{Insert}, \text{Delete}\}$*/
    $m \leftarrow m \pm 1$;  /*Insert: $+1$, Delete: $-1$*/
    **for** *each* $i \in [0, L]$:
        $c_q^i \leftarrow$ the center of the cell contains $q$ at level $i$;
        $\text{HH}_i$.update$(op, c_q^i)$;
        **for** *each* $o \in [O]$:
            **if** $h_{o,i}(q) == 1$:
                $\text{SC}_{o,i}$.update$(op, c_q^i)$;

**Query**:

Let $o^*$ be the smallest $o$ such that no instance of $\text{SC}_{o,0}, \text{SC}_{o,1}, \ldots, \text{SC}_{o,L}$ returns Fail;
$S \leftarrow \{\}$;
$\mathcal{C}_{-1} \leftarrow$ the cell of the entire space $[\Delta]^d$; $\widehat{|\mathcal{C}_{-1}|} \leftarrow m$;
**for** $i \in [0, L-1]$:
    $C_i \leftarrow \text{HH}_i$.query().top$((e+4)(L+1)k/\rho)$;
    Remove cells $\mathcal{C}$ from $C_i$ if $\mathcal{C}^P(\mathcal{C}) \notin C_{i-1}$, where $\mathcal{C}^P(\mathcal{C})$ is the parent cell of $\mathcal{C}$ in level $i-1$;
    $B_i \leftarrow \text{SC}_{o^*, i}$.query();
    $S_i \leftarrow \{p \in B_i : \mathcal{C}(p, i-1) \in C_{i-1} \text{ AND } \mathcal{C}(p, i) \notin C_i\}$;
    Each point in $\mathcal{S}_i$ receives weight $1/\pi_i(o^*)$;
    $S \leftarrow S \cup S_i$;
$k' \leftarrow \sum_{i \in [0, L]} |C_i|$;
Let $\{\mathcal{C}_1, \mathcal{C}_2, \ldots \mathcal{C}_{k'}\} = \cup_{i \in [0, L]} C_i \cup \{\mathcal{C}_{-1}\}$ be the set of heavy cells;
Let $\{\widehat{|\mathcal{C}_1|}, \widehat{|\mathcal{C}_2|}, \ldots \widehat{|\mathcal{C}_{k'}|}\}$ be the estimated frequency of each cell;
$R \leftarrow \text{RectifyWeights}(\widehat{|\mathcal{C}_1|}, \widehat{|\mathcal{C}_2|}, \ldots \widehat{|\mathcal{C}_{k'}|}, S)$;
**return** $R$.